\documentclass[preprint2]{aastex631}

\usepackage{graphicx} 
\def\approxgt{\ifmmode \rlap{$>$}{}_{{}_{{}_{\textstyle\sim}}} \else%
$\rlap{$>$}{}_{{}_{{}_{\textstyle\sim}}}$\fi} 
\def\flux{erg cm$^{-2}$ s$^{-1}$}

\def\arcsec{\hbox{$^{\prime\prime}$}}
\def\approxlt{\ifmmode \rlap{$<$}{}_{{}_{{}_{\textstyle\sim}}} \else%
$\rlap{$<$}{}_{{}_{{}_{\textstyle\sim}}}$\fi}
\begin{document}

\title{The Einstein Probe transient EP240414a: Linking Fast X-ray Transients, \\ Gamma-ray Bursts and Luminous Fast Blue Optical Transients}


\author{Joyce N. D. van Dalen}
\affiliation{{Department of Astrophysics/IMAPP}, {Radboud University Nijmegen}, {{P.O.~Box 9010}, {Nijmegen}, {6500~GL},   {The Netherlands}}}

\author{Andrew J. Levan}
\affiliation{{Department of Astrophysics/IMAPP}, {Radboud University Nijmegen}, {{P.O.~Box 9010}, {Nijmegen}, {6500~GL},   {The Netherlands}}}
\affiliation{{Department of Physics}, {University of Warwick}, {{Coventry, CV4 7AL}, {UK}}}

\author{Peter G. Jonker}
\affiliation{{Department of Astrophysics/IMAPP}, {Radboud University Nijmegen}, {{P.O.~Box 9010}, {Nijmegen}, {6500~GL},   {The Netherlands}}}
\affiliation{{SRON}, {Netherlands Institute for Space Research}, {{Niels Bohrweg 4}, {Leiden, 2333 CA}, {The Netherlands}}}

\author{Daniele Bj\o{}rn Malesani}
\affiliation{{Cosmic Dawn Center (DAWN)}, {{Denmark}}}
\affiliation{{Niels Bohr Institute}, {University of Copenhagen}, {{Jagtvej 128}, {Copenhagen}, {2200}, {Denmark}}}

\author{Luca Izzo}
\affiliation{{Osservatorio Astronomico di Capodimonte}, {INAF}, {{Salita Moiariello 16}, {Napoli}, {80131},   {Italy}}}
\affiliation{{Niels Bohr Institute, University of Copenhagen}, {DARK}, {{Jagtvej 128}, {Copenhagen}, {2200},   {Denmark}}}

\author{Nikhil Sarin}
\affiliation{{The Oskar Klein Centre, Department of Physics}, {Stockholm University}, {{AlbaNova}, {Stockholm}, {SE-106 91}, {Stockholm}, {Sweden}}}
\affiliation{{Nordita}, {Stockholm University and KTH Royal Institute of Technology}, {{Hannes Alfvéns väg 12}, {Stockholm}, {SE-106 91}, {Stockholm}, {Sweden}}}

\author{Jonathan Quirola-V\'asquez}
\affiliation{{Department of Astrophysics/IMAPP}, {Radboud University Nijmegen}, {{P.O.~Box 9010}, {Nijmegen}, {6500~GL},   {The Netherlands}}}

\author{Daniel Mata S\'anchez}
\affiliation{ Instituto de Astrofı\'isica de Canarias, IAC, E-38205, La Laguna, Tenerife, Spain}
\affiliation{{Departamento de Astrof\'isica, Univ. de La Laguna},   {{E-38206}, {La Laguna, Tenerife},     {Spain}}}

\author{Antonio de Ugarte Postigo}
\affiliation{{Universit\'{e} de la C\^ote d'Azur, Observatoire de la C\^ote d'Azur, CNRS, Artemis},   {  {Nice}, {F-06304},   {France}}}
\affiliation{{Aix Marseille Univ, CNRS, LAM}, {{Marseille},     {France}}}

\author{Agnes P. C. van Hoof}
\affiliation{{Department of Astrophysics/IMAPP}, {Radboud University Nijmegen}, {{P.O.~Box 9010}, {Nijmegen}, {6500~GL},   {The Netherlands}}}

\author{Manuel A. P. Torres}
\affiliation{ Instituto de Astrofı\'isica de Canarias, IAC, E-38205, La Laguna, Tenerife, Spain}
\affiliation{{Departamento de Astrof\'isica, Univ. de La Laguna},   {{E-38206}, {La Laguna, Tenerife},     {Spain}}}

\author{Steve Schulze}
\affiliation{Center for Interdisciplinary Exploration and Research in Astrophysics (CIERA), Northwestern University, 1800 Sherman Ave, Evanston, IL 60201, USA}

\author{Stuart P. Littlefair}
\affiliation{{Department of Physics and Astronomy}, {University of Sheffield}, {  {Sheffield}, {S3 7RH},   {United Kingdom}}}

\author{Ashley Chrimes}
\affiliation{{European Space Agency (ESA)}, {European Space Research and Technology Centre (ESTEC)}, {{Keplerlaan 1}, {Noordwijk}, {2201~AZ},   {The Netherlands}}}
\affiliation{{Department of Astrophysics/IMAPP}, {Radboud University Nijmegen}, {{P.O.~Box 9010}, {Nijmegen}, {6500~GL},   {The Netherlands}}}

\author{Maria E. Ravasio}
\affiliation{{Department of Astrophysics/IMAPP}, {Radboud University Nijmegen}, {{P.O.~Box 9010}, {Nijmegen}, {6500~GL},   {The Netherlands}}}
\affiliation{Osservatorio Astronomico di Brera, Istituto Nazionale di Astrofisica, Merate 23807, Italy}

\author{Franz E. Bauer}
\affiliation{{Instituto de Astrof{\'{\i}}sica, Facultad de F{\'{i}}sica and Centro de Astroingenier{\'{\i}}a, Facultad de F{\'{i}}sica, Pontificia Universidad Cat{\'{o}}lica de Chile}, {UC}, {{Campus San Joaquín, Av. Vicuña Mackenna 4860, Macul}, {Santiago}, {7820436}, {RM}, {Chile}}}
\affiliation{{Millennium Institute of Astrophysic}, {MAS}, {{Nuncio Monse{\~{n}}or S{\'{o}}tero Sanz 100, Of 104, Providencia}, {Santiago}, {7500011}, {RM}, {Chile}}}
\affiliation{{Space Science Institute}, {SSI}, {{4750 Walnut Street, Suite 205}, {Boulder}, {80301}, {CO}, {USA}}}

\author{{Antonio} {Martin-Carrillo}}
\affiliation{{School of Physics and Centre for Space Research}, {University College Dublin}, {{Belfield}, {Dublin 4},     {Ireland}}}

\author{Morgan Fraser}
\affiliation{{School of Physics and Centre for Space Research}, {University College Dublin}, {{Belfield}, {Dublin 4},     {Ireland}}}


\author{Alexander J. van der Horst}
\affiliation{{Department of Physics}, {George Washington University}, {725 21st St NW}, {Washington}, {DC}, {20052}, {USA}}

\author{{Pall} {Jakobsson}}
\affiliation{{Centre for Astrophysics and Cosmology}, {University of Iceland}, {{Dunhagi}, {Reykjav\'ik}, {107},   {Iceland}}}

\author{{Paul} {O'Brien}}
\affiliation{{School of Physics and Astronomy}, {University of Leicester}, {{University Road}, {Leicester}, {LE1 7RH},   {United Kingdom}}}

\author{{Massimiliano} {De Pasquale}}
\affiliation{{MIFT Department}, {University of Messina}, {{Via F. S. D'Alcontres 31}, {Messina}, {98166},   {Italy}}}

\author{{Giovanna} {Pugliese}}
\affiliation{{Anton Pannekoek Institute of Astronomy}, {University of Amsterdam}, {{Science Park 904}, {Amsterdam}, {1098 XH},   {The Netherlands}}}

\author{Jesper Sollerman}
\affiliation{{The Oskar Klein Centre, Department of Physics}, {Stockholm University}, {{AlbaNova}, {Stockholm}, {SE-106 91}, {Stockholm}, {Sweden}}}

\author{Nial R. Tanvir}
\affiliation{{School of Physics and Astronomy}, {University of Leicester}, {{University Road}, {Leicester}, {LE1 7RH},   {United Kingdom}}}


\author{{Tayyaba} {Zafar}}
\affiliation{{School of Mathematical and Physical Sciences}, {Macquarie University}, {    {2109}, {NSW}, {Australia}}}


\author{Joseph P. Anderson}
\affiliation{European Southern Observatory, Alonso de C\'ordova 3107, Casilla 19, Santiago, Chile}
\affiliation{{Millennium Institute of Astrophysic}, {MAS}, {{Nuncio Monse{\~{n}}or S{\'{o}}tero Sanz 100, Of 104, Providencia}, {Santiago}, {7500011}, {RM}, {Chile}}}

\author{Llu\'is Galbany}
\affiliation{Institute of Space Sciences (ICE, CSIC), Campus UAB, Carrer de Can Magrans, s/n, E-08193 Barcelona, Spain.}
\affiliation{Institut d'Estudis Espacials de Catalunya (IEEC), 08860 Castelldefels (Barcelona), Spain.}

\author{Avishay Gal-Yam}
\affiliation{Department of Particle Physics and Astrophysics, Weizmann Institute of Science, 234 Herzl St, Rehovot, 76100, Israel.}

\author{Mariusz Gromadzki}
\affiliation{Astronomical Observatory, University of Warsaw, Al. Ujazdowskie 4, 00-478 Warszaw, Poland}

\author{Tomás E. Müller-Bravo}
\affiliation{Institute of Space Sciences (ICE, CSIC), Campus UAB, Carrer de Can Magrans, s/n, E-08193 Barcelona, Spain.}
\affiliation{Institut d'Estudis Espacials de Catalunya (IEEC), 08860 Castelldefels (Barcelona), Spain.}

\author{Fabio Ragosta}
\affiliation{Dipartimento di Fisica “Ettore Pancini”, Università di Napoli Federico II, Via Cinthia 9, 80126 Naples, Italy }
\affiliation{INAF - Osservatorio Astronomico di Capodimonte, Via Moiariello 16, I-80131 Naples, Italy}

\author{Jacco H. Terwel}
\affiliation{School of Physics, Trinity College Dublin, The University of Dublin, Dublin 2, Ireland}
\affiliation{Nordic Optical Telescope, Rambla José Ana Fernández Pérez 7, ES-38711 Breña Baja, Spain}


\begin{abstract}

    Detections of fast X-ray transients (FXTs) have been accrued over the last few decades. However, their origin has remained mysterious. There is now rapid progress thanks to timely discoveries and localisations with the Einstein Probe mission. Early results indicate that FXTs may frequently, but not always, be associated with gamma-ray bursts (GRBs). Here, we report on the multi-wavelength counterpart of FXT EP240414a, which has no reported gamma-ray counterpart. The transient is located 25.7~kpc in projection from a massive galaxy at $z=0.401$. We perform comprehensive photometric and spectroscopic follow-up. The optical light curve shows at least three distinct emission episodes with timescales of $\sim 1, 4$ and 15 days and peak absolute magnitudes of $M_R \sim -20$, $-21$, and $-19.5$, respectively. The optical spectrum at early times is extremely blue, inconsistent with afterglow emission. It may arise from the interaction of both jet and supernova shock waves with the stellar envelope and a dense circumstellar medium, as has been suggested for some Fast Blue Optical Transients (LFBOTs). At late times, the spectrum evolves to a broad-lined~Type~Ic supernova, similar to those seen in collapsar long-GRBs. This implies that the progenitor of EP240414a is a massive star creating a jet-forming supernova inside a dense envelope, resulting in an X-ray outburst with a luminosity of $\sim 10^{48}$ erg s$^{-1}$, and the complex observed optical/IR light curves. If correct, this argues for a causal link between the progenitors of long-GRBs, FXTs and LFBOTs.
    
\end{abstract}


\section{Introduction}
Short-duration astrophysical transients span the electromagnetic spectrum from radio to optical, X-rays, and gamma-rays. These events take place on timescales of milliseconds in the case of Fast Radio Burst (FRBs) \citep[e.g.,][]{Petroff2016, Petroff2019, Cordes2019}, fractions of a second to hours for gamma-ray bursts (GRBs) \citep[e.g.,][]{Kouveliotou93,Ackermann2013, levan14, Ajello2019}, and hours to days for kilonovae (KNe) \citep[e.g.,][]{tanvir13,2017Abbott,Metzger19} and days to months for supernovae (SNe) \citep[e.g.,][]{Kann2011, Prentice2018, Ho2023}. The rapid multi-wavelength follow-up they often receive is necessary to uncover their nature. 

The origins of fast X-ray transients (FXTs), which are bursts of soft X-ray photons ($\approx0$.3--10 keV) have so far been amongst the most elusive. Although FXTs have been identified in X-ray monitors from the early days of sounding rockets \citep[e.g.,][]{cooke76,rappaport76}, their current samples have been mostly built through intensive searches of {\em Chandra} and {\em XMM-Newton} archival data \citep[e.g.,][]{Jonker2013, Glennie2015, Bauer2017, Alp2020, 2020Novara, Lin2022, Quirola-Vasquez2022, Quirola-Vasquez2023}. Over thirty bursts lasting from hundreds to thousands of seconds with power-law spectral shapes have been discovered. The origin of these bursts is yet unclear. Proposed explanations include: the formation of the burst in the events following a binary neutron star merger which leads to the formation of a millisecond pulsar whose rapid spin down powers the burst \citep[e.g.,][]{Zhang2013, Metzger2014}; a tidal disruption event (TDE) involving a white dwarf disrupted by an intermediate-mass black hole \citep[e.g.,][]{Jonker2013, Macleod2016, Bauer2017}; a shock breakout after a compact progenitor SN \citep[e.g.,][]{Soderberg2008, Waxman2017}; or a jet breakout following long-GRBs through cocoon-like emission \citep[e.g.,][]{Nakar2017, Izzo2019}. The discovery of FXTs predominantly in archival data - long after the events - has complicated the distinction between these different origins as possible multi-wavelength counterparts have not been observed except in the case of SN 2008D \citep{Soderberg2008} and recently in EP240315a \citep{EP_paper,Levan2024, 2024Gillanders}. Aside from these events, most insight has been provided via observations of the likely host galaxies, whose redshifts, in turn, enable the energetics of the event to be determined (see e.g., \citealt{2022Eappachen,Quirola-Vasquez2022, Quirola-Vasquez2023, 2024Inkenhaag}).

The Einstein Probe (EP) mission \citep{Yuan2015, Yuan2022}, which was launched on the January 9 2024, was designed to search for and follow-up high energy transients. The  Wide-field X-ray Telescope (EP-WXT), with its large 3600 deg$^{2}$ field of view is sensitive in the 0.5–4 keV band, and now provides timely alerts of new transients. This presents an opportunity to follow-up a large number of FXTs shortly after their detection. Multiple such sources have now been announced, of which EP240315a was the first FXT with an observed optical and radio counterpart \citep{EP_paper,Levan2024, 2024Gillanders}. That distant event (\textit{z}=4.859) was associated with a long-duration $\gamma$-ray burst, and it has been proposed that GRBs of varying luminosities could explain a significant portion of the FXT population \citep{Levan2024}. Several more FXTs have been reported \citep[e.g.,][]{EP240219a,EP240617a,EP240703a,EP240801a,EP240807a,EP240913a,EP240919a}, some coincident with GRBs, supporting a scenario in which GRBs and FXTs are linked. However, it is also striking that not all FXTs have associated GRBs, even when $\gamma$-ray telescopes were sensitive to their detection. This raises the question of whether the majority of the extragalactic FXT population is related to GRBs, or if more complex progenitor scenarios should be considered.

Here we consider the case of EP240414a, the second identified EP source to have a secure multi-wavelength counterpart, and an example in which no coincident GRB was reported. We present extensive imaging and spectroscopic observations spanning from 0.5-100 days after the EP-WXT trigger and show that these observations support a link to lon-GRB progenitors, but reveal a complex counterpart behaviour that does not mirror that observed for most long-GRBs. Throughout we provide magnitudes in the AB system, and assume a $\Lambda$CDM cosmology with $H_0=70.0$ km s$^{-1}$ Mpc$^{-1}$ and $\Omega_{\Lambda}$=0.68.

\begin{figure*}[ht!]
    \centering
    \includegraphics[width=\linewidth]{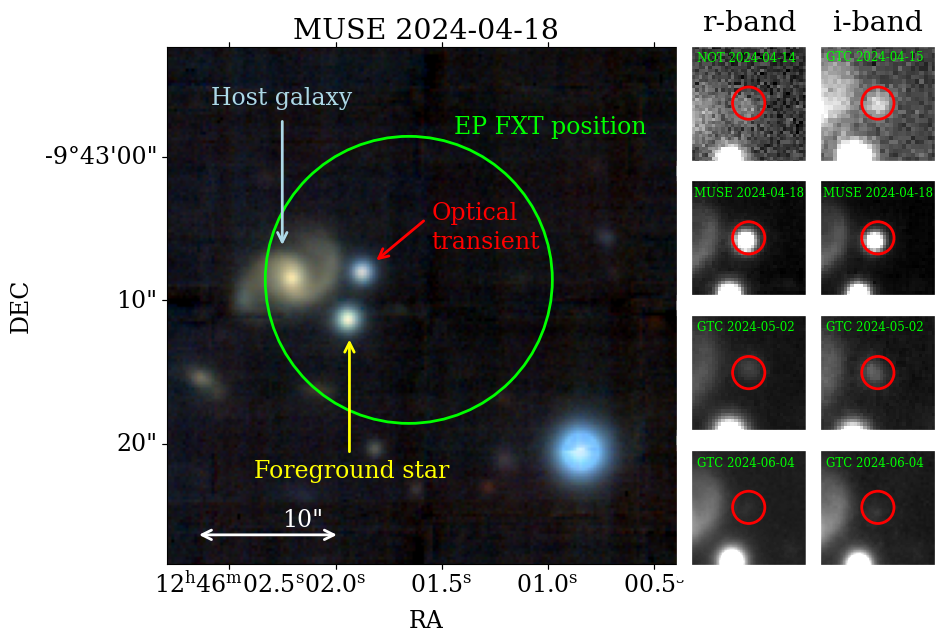}
    \caption{\textit{Large panel}: false-color RGB image from the MUSE observation in the Johnson V and the Cousins R and I bands. The location of the transient as reported by the EP FXT instrument is given in green with by the 10\arcsec~radius 90~$\%$ C.L. \citep{Guan2024}. The optical transient marked with OT, the host galaxy and a foreground star within the 90~$\%$ C.L. confidence interval are marked with red, blue and yellow respectively. \textit{Small panels}: \textit{r}- and \textit{i}-band imaging at 4 different epochs. }
    \label{fig:imaging}
\end{figure*} 

\section{Observations}\label{observations}

EP240414a was detected by the EP WXT on 2024-04-14 09:50:12 UTC \citep{Lian2024}, who reported a peak flux of $~3\times10^{-9}$ erg s$^{-1}$ cm$^{-2}$ in the 0.5-4 keV band. Observations taken from Lulin Observatory, Taiwan at $\sim 3.13$ hours after the initial detection revealed a new optical source within the 3 arcminute localisation area \citep{Aryan2024}. This was suggested as the optical counterpart of EP240414a (the optical counterpart was named AT2024gsa), lying close to the cataloged SDSS galaxy with an active galactic nucleus (AGN) tabulated as \textit{SDSS J124601.99-094309.3} \citep{SDSS}. The identifications became more secure when an X-ray location for the fading transient light was reported from the EP Follow-up X-ray Telescope \citep{Guan2024}. Our observations began with imaging and spectroscopy from La Palma $\sim 12$ hours after the EP-WXT trigger, providing a redshift of \textit{z}$\approx$0.401 for the host galaxy \citep{Jonker20242}. Further observations over the following nights showed unexpectedly that rather than a continuous fading the counterpart brightened markedly between 2-3 days after discovery, very different from the behaviour seen in most other high-energy transients. A bright radio counterpart was also discovered on 23 April \citep{Bright2024}. However, no detections or upper limits in gamma-rays have been reported from the major gamma-ray satellites. We note that the {\em Fermi} satellite was in the South Atlantic Anomaly (SAA) at the time of the EP WXT detection, and no other high-energy satellites have reported either gamma-ray limits or detections.

Below we describe our observational campaign of EP240414a but also refer the reader to \citet[2024,][]{Bright0414a,Shubham0414a, EP0414a} for additional observations in the radio, optical, and X-ray regimes respectively. In this work, we refer to EP240414a by name or as a burst (for clarity we will not use AT2024gsa specifically for the optical transient). Logs of our photometry, spectroscopy and X-ray data are shown in Tables~\ref{tab:photometry}-\ref{tab:host}. All our photometry is aperture photometry aligned and calibrated to Pan-STARRS and 2MASS stars in the field. The resulting magnitudes are given in the AB system. All our spectra are wavelength- and flux-calibrated, and cosmic-ray-corrected.

\subsection{Gran Telescopio Canarias (GTC) spectroscopy and
photometry}
The Gran Telescopio Canarias (GTC) at the Roque de los Muchachos observatory (Canary Islands, Spain) took observations of the source using the OSIRIS+, HiPERCAM, and EMIR instruments on several nights (program GTC1-ITP23 PIs Jonker, Torres).

\subsubsection{OSIRIS}
Two epochs of spectroscopy were obtained with the Optical System for Imaging and low-intermediate-Resolution Integrated Spectroscopy (OSIRIS) instrument \citep{2000Cepa}. The first spectrum of the source was obtained on 15~April 2024 ($\sim$0.6~days since trigger) by OSIRIS+ with grism R1000R in a 1200~s exposure with the slit placed at a position angle of $-$82 degrees to cover both the candidate host galaxy and transient location. Further spectroscopy was obtained on the 3~May ($\sim$20~days since trigger) in grism R1000R for an exposure time of 1200~s with the final spectrum oriented at $-$16.7 degrees to minimize galaxy background and place a foreground star on the slit. The data reduction for the OSIRIS+ spectra included bias subtraction and flat-field correction according to default \texttt{PyRAF} tasks \citep{Pyraf2012}, and cosmic rays correction with the \texttt{LACosmic} task \citep{vanDokkum2001}. We note that the flux calibration of the first observation was performed using a standard star observed at the end of the night and therefore there is an intrinsic uncertainty on the calibration. We nevertheless made an effort to correct for slit losses, considering wavelength dependent seeing and airmass correction to obtain a better calibration. We list these spectroscopic observations in Table~\ref{tab:spectroscopy}.

\subsubsection{HiPERCAM}
Photometry in the \textit{ugriz}-bands was obtained by HiPERCAM \citep{2016Dhillon, 2018Dhillon, 2021Dhillon} on the nights of 2~May ($\sim$19~days since trigger) and 4~June ($\sim$51~days since trigger) in 16$\times$60~s exposures, and data were reduced with the HiPERCAM pipeline. The source is detected in the \textit{griz} bands and the photometry is provided in Table~\ref{tab:photometry}. The \textit{r}- and \textit{i}-band images are shown in Figure~\ref{fig:imaging}.

\subsubsection{EMIR}
Near-infrared photometry of the late-time emission was obtained with EMIR~\citep{2022Garzon} on 17~May ($\sim$34~days since trigger). The target was observed in the $K_{s}$ and \textit{J} bands for 1440~s and 1400~s respectively. The EMIR data were reduced using a custom pipeline to create flat field and sky frames and correct astrometry. 

\subsection{Very Large Telescope (VLT) spectroscopy and photometry}
The Very Large Telescope (VLT) at Cerro Paranal, Chile observed the source with FORS2, MUSE, and X-shooter on multiple nights under different programs. Logs of the photometric and spectroscopic observations are provided in Table~\ref{tab:photometry} and Table~\ref{tab:spectroscopy}, respectively. 

\subsubsection{FORS2}
The FORS2 instrument \citep{1998Appenzeller} was used to obtain spectroscopy in 300V grism (113.26ET.002, PI Jonker) and imaging in $g, r, i$ and $z$ (113.26ET.008, PI Jonker) on 19~April ($\sim$5~days since trigger) and 8~May ($\sim$24~days since trigger) respectively. We reduced the spectroscopic observations using standard procedures in IRAF~\citep{1986IRAF}. We flux-calibrated the spectrum using observations of the spectrophotometric standard EG 274 obtained with the same setup immediately following the science observations. Flux calibration appears less reliable on the blue end of the spectrum resulting in a broad bump below approximately 4000~\AA~(See Fig. \ref{fig:SED}). This broad feature does not imprint on the narrow lines in the spectrum, yet the region should be treated with some caution.

The imaging in $g, r, i$ and $z$ were taken at 3$\times$100~s, 3$\times$50~s, 3$\times$50~s, and 5$\times$40~s exposures respectively. Individual images were reduced with the ESO FORS2 image data reduction pipeline and combined with standard tools in PyRAF~\citep{Pyraf2012}.

\subsubsection{MUSE}
We obtained two epochs of observations with the MUSE instrument \citep{2010Bacon}. The first epoch was taken on 18~April at $\sim$4~days since trigger (ESO 110.24CF.022 1110.A-4348, PI Tanvir), and the second with an identical set-up on 4 July at $\sim$81~days since trigger (ESO 111.259Q.001, PI Jonker). The data reduction for both epochs was done in ESOREX using the MUSE data reduction pipeline \citep{ESOREX2015, MUSE2020} and additional sky subtraction was performed with Zurich Atmosphere Purge (ZAP) \citep{ZAP2015}. 
From both epochs, we extract images in the \textit{i}-, \textit{r}-, and \textit{z}-band and extract spectra at the transient location with the Python package MPDAF  \citep{MPDAF2016}. The first MUSE observation was taken near the peak magnitude of the transient at $\sim$4~days. Figure~\ref{fig:imaging} shows a color composite image of this observation in Johnson V and the
Cousins R and I bands, and the extracted \textit{r}- and \textit{i}-band images at this epoch. 

\subsubsection{X-shooter}
X-shooter \citep{2011Vernet} observed EP240414a on 25 April ($\sim$11~days since trigger) with single 1200~s exposures in the UVB and VIS arm and 6$\times$300s exposures in the NIR arm in nodding mode (ESO 113.26ET.004, PI Jonker). Seeing was between 0.45\arcsec~and 0.69\arcsec~during these observations at an airmass between 1.04 and 1.08. The data was reduced with the ESO X-shooter pipeline \citep{XSHOOTER2010}.

\subsection{New Technology Telescope (NTT) photometry}
The advanced Public ESO Spectroscopic Survey for Transient Objects (ePESSTO+; \citealt{2015PESSTO}) observed the source on 15 April ($\sim$2~days since trigger) with the New Technology Telescope (NTT) at the La Silla Observatory (Chile). Photometry in \textit{gri}-bands was taken by the EFOSC2 instrument \citep{1984Buzzoni} for 3x250~s as described in Table~\ref{tab:photometry}. Data reduction of the images was done with the ESO EFOSC pipeline \citep{2010Izzo}. The transient was detected in \textit{g}-band, and upper-limits in \textit{r}- and \textit{i}-band were obtained. 

\subsection{Nordic Optical Telescope (NOT) photometry}
The Nordic Optical Telescope (NOT) at the Roque de los Muchachos observatory (Canary Islands, Spain) observed the source on multiple nights with the ALFOSC camera in the $g, r, i$ and $z$-bands. 
We use a standard dithering pattern for all observations. The exposure times for each of the observations are noted in Table~\ref{tab:photometry}. The images were reduced with the standard data reduction steps in IRAF~\citep{1986IRAF}.

\subsection{Southern Astrophysical Research Telescope (SOAR) photometry}
Follow-up using the Goodman spectrograph at the SOAR 4.1m telescope \citep{Clemens2004}, and the Red Camera instrument in imaging mode was done at two different epochs (program SOAR2024A-012, Bauer PI). On 28~April at $\sim$14~days since trigger and 28~May at $\sim$44~days since trigger $i$-band images were obtained. The data were bias subtracted and flat-field corrected adopting standard \texttt{PyRAF} tasks \citep{Pyraf2012}, and cosmic rays corrected using the \texttt{LACosmic} task \citep{vanDokkum2001}. The source was weakly detected at the first epoch, but not in the second and the photometric results are provided in Table~\ref{tab:photometry}.

\subsection{Swift}

\textit{Swift} observed the location of EP240414a responding to three different target-of-opportunity requests (PIs Liu, Evans, and Levan). The data, which are public, span the time range between 2024 April~18 and June~4, for a total exposure time of about 19.3~ks, split in eight visits. This provides us with X-ray data between approximately 4 and 22 days.

The 0.3--10 keV count-rate light curve was retrieved using the public automated online tool provided by the University of Leicester \citep{2007A&A...469..379E,2009MNRAS.397.1177E}. Significant detection was achieved only at two epochs, while only upper limits could be secured afterward. Besides, the few accumulated counts were not sufficient to extract a well-constrained spectrum. To convert the light curve to physical units, a typical afterglow spectrum was assumed with photon index $\Gamma = 2$ and an absorbing column density equal to the Galactic value $N_{\rm H} = 3.3\times10^{20}$~cm$^{-2}$ \citep[e.g.,][]{2013Willingale}. The resulting flux values are listed in Table~\ref{tab:x-rays}, with the exception of late-time upper limits, which were not competitive compared to the nearly simultaneous \textit{Chandra} observations.

In addition, we analysed the images obtained by the {\em Swift}-UVOT in the UVM2 filter to obtain both a weak detection of the transient emission at $\sim 4$ days after the burst, and a late time detection of the host galaxy.

\begin{figure*}
    \centering
    \includegraphics[width=\textwidth]{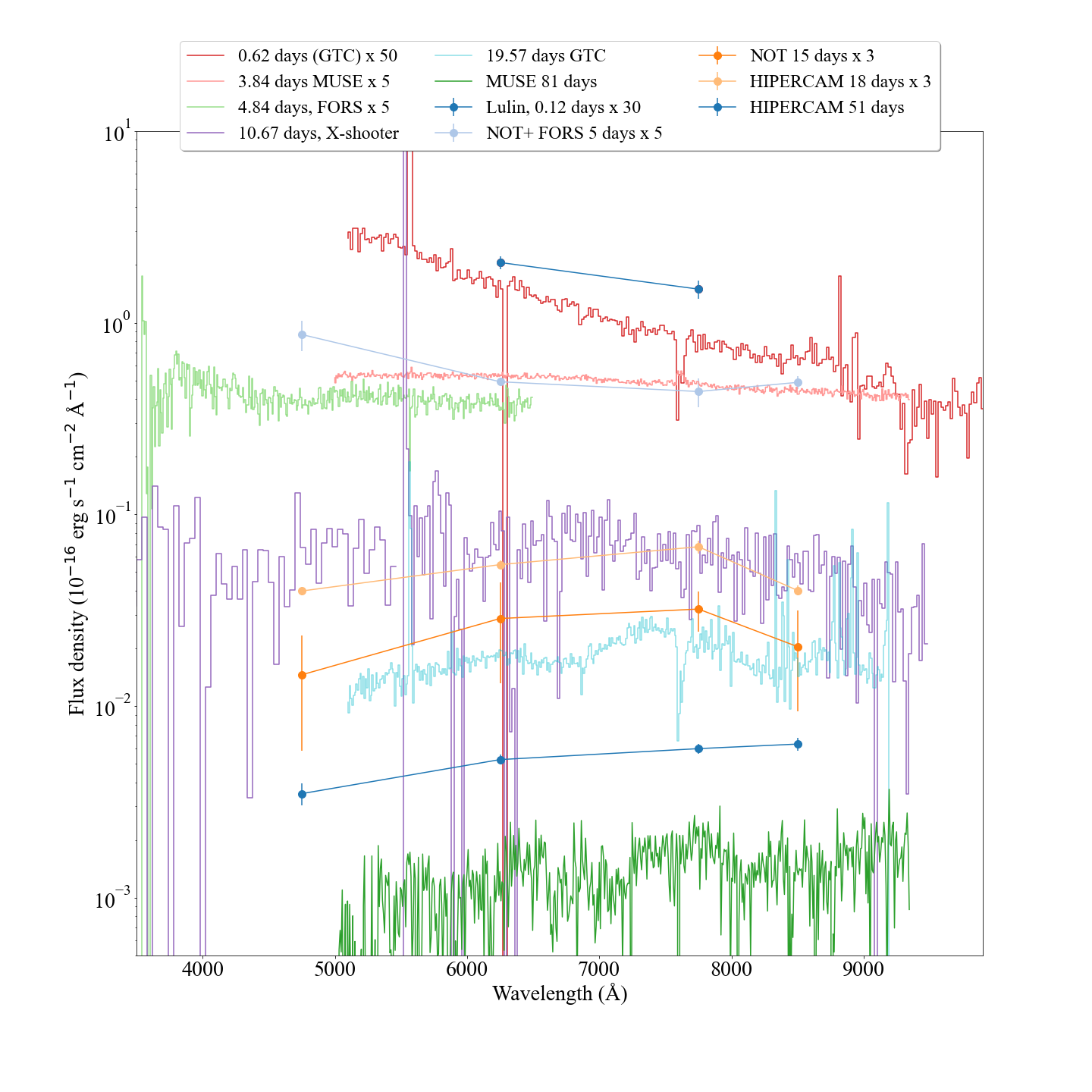}
    \caption{The evolution of the spectrum and spectral energy distribution of EP24014a in the observer reference frame. There is strong evolution from a very blue early spectral slope, through a near featureless flat spectrum to a thermal spectrum with additional broad features of a SN~Ic-BL.}
    \label{fig:SED}
\end{figure*}

\subsection{Chandra}
{\em Chandra} observed the location of EP~240414a on June 16 ($\sim$62~days since trigger) for 10.7~ks (Director's Discretionary Time). The source position was placed at the S3 CCD of the ACIS-S detector array using the very faint mode (\citealt{1997AAS...190.3404G}). We analyzed the data using the data analysis package {\sc ciao}~version 4.16 (Chandra Interactive Analysis of Observations) \citep{2006Ciao}.
Using {\sc wavdetect} in {\sc ciao} to detect sources in the observation, we detect X-ray emission spatially co-incident with the nucleus of the host galaxy of EP~240414a (\citealt{2024GCN.36362....1B}), and consistent with the presence of broad-lines in the optical spectrum (see Section~\ref{host}). We extract a spectrum via {\sc ciao} and fit this with the detector-independent spectral fitting program \texttt{XSPEC} (v12.14) \citep{1996Arnaud} using a power-law plus Galactic absorption model ({\sc tbabs},  N$_H=3.35\times 10^{20}$~cm$^{-2}$). The best-fit power law index is 1.1$\pm$0.9 ($\Delta C=2.7$) and the best-fit source absorbed flux $F_X$=$(5^{+4}_{-2})\times 10^{-14}$ \flux\ (0.5-10 keV) for a C-stat of 13.8 for 16 degrees of freedom \citep{1979Cash}. We note that the point spread function of the EP-FXT and {\em Swift} XRT contains light from both the AGN in the host galaxy and from the counterpart and so, on the assumption that the AGN is not significantly varying, this flux is be subtracted from those data to obtain transient only flux in this work.

At the location of the transient, we find two photons, which corresponds to a 95\% confidence upper limit of about 4.7 photons using the statistics of \citet{1991ApJ...374..344K} and an estimate of the background at the position of the source. Assuming a photon index of 2 and Galactic absorption as above, the corresponding 0.5-10 keV flux upper limit is $F_X <8.5\times 10^{-15}$ erg cm$^{-2}$ s$^{-1}$ at $2 \sigma$. We give the AGN flux and transient flux in Table~\ref{tab:x-rays}.

\section{Results}\label{results}

\subsection{Redshift and Energetics}

Our spectroscopic observations of EP240414a and its host reveal a clear redshift for the host galaxy at \textit{z}=0.401, which we obtain from a fit to the H$\alpha$ emission line (see Section \ref{host} for the analysis of the host galaxy). 
While we do not detect strong absorption lines in the early spectra of the transient, we can securely tie it to the host since the later spectroscopy reveals clear features of a broad-lined~Type~Ic~SN (SN~Ic-BL) at $z=0.401$ (see Section~\ref{spectroscopy} and Figure~\ref{fig:SED}). Hence, we conclude that EP240414a lies at $z=0.401$ and is associated with the host galaxy \textit{SDSS J124601.99-094309.3} at the same redshift. 

\begin{figure*}[ht!]
    \centering
    \includegraphics[width=\linewidth]{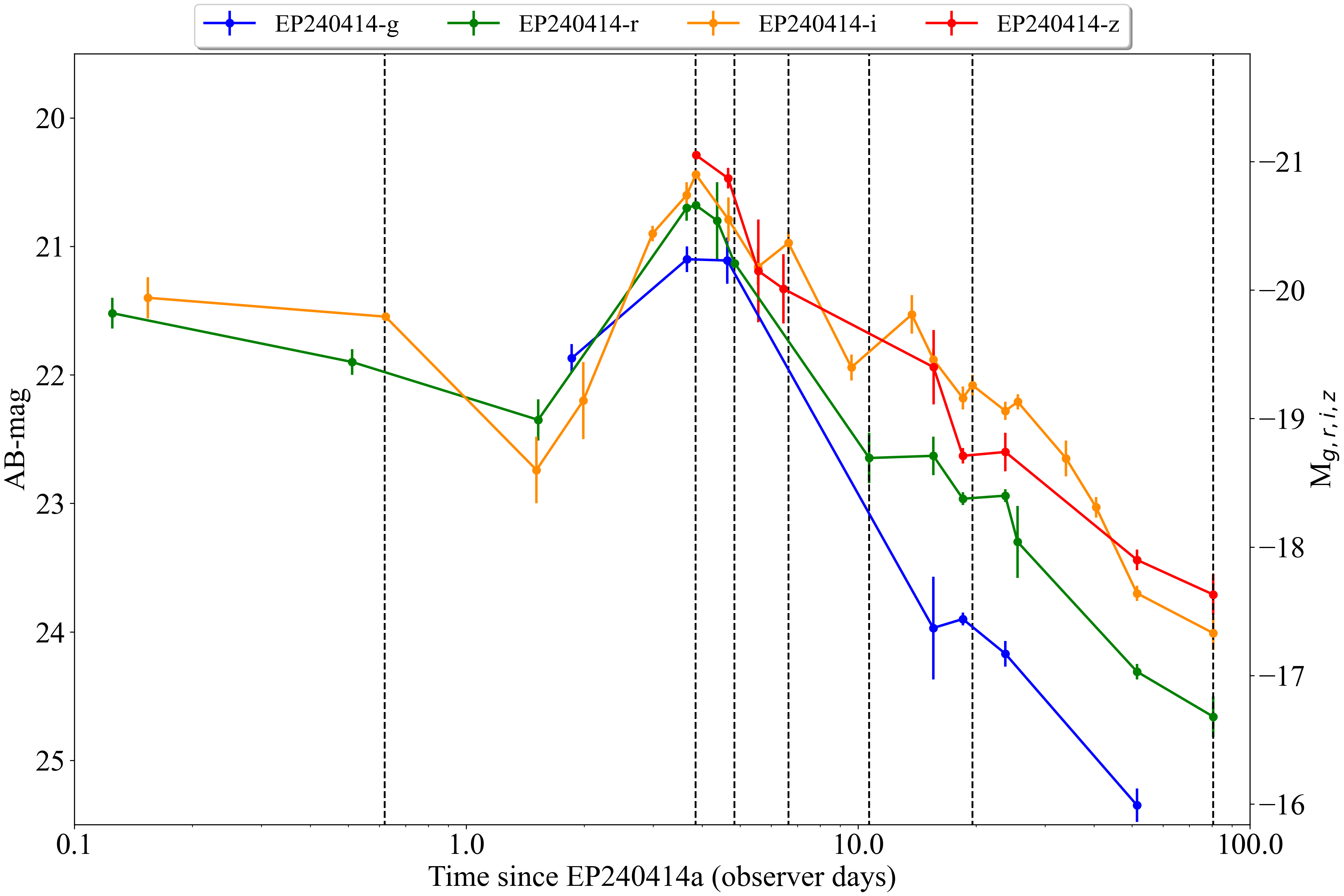}
    \caption{Light curve of EP240414a in the observer frame in \textit{g}- (blue), \textit{r}- (green), \textit{i}- (yellow) and \textit{z}-band (red). The epochs at which our spectroscopic observations were taken are indicated by the black dashed lines. The \textit{r}- and \textit{i}-band light curve shows moderate fading within the first day which we call the first peak. Then, we see rebrightening between day 2 and 3 in \textit{g}-, \textit{r}-, and \textit{i}-band which is followed by rapid fading after 4 days in all bands, to which we refer as the second peak. Modest rebrightening in \textit{i}-band and flattening of the slope in the other bands at $\sim 10$ days is observed consistent with spectroscopic observations of the SN which is referred to as the third peak.}
    \label{fig:lc}
\end{figure*}

At this redshift the X-ray outburst has a peak luminosity of $L_{X,iso} \approx 2 \times 10^{48}$ erg s$^{-1}$ in the 0.5-4 keV band \citep{Lian2024}. Although this luminosity is neither bolometric, nor in the band typically used for GRBs, it is a luminosity substantially fainter than implied for the vast majority of GRBs and indicates a low luminosity event \citep{2022Dainotti}. The bolometric correction is highly uncertain since it depends on the spectrum of the emission well beyond of the observed band. For example for a typical GRB spectral shape , with corrections from 0.5-4 keV to the typical 1-10000 keV band used for energetics, the bolometric correction varies from $\sim$ 3 for $E_p = 10$ keV to $>50$ for $E_p > 100$ keV
(assuming a Band spectrum [\citealt{1993Band}] with $\alpha = -1.0$, $\beta = -2.3$). However, we might have expected events with much higher $E_p$ to trigger $\gamma$-ray detectors with all sky sensitivity such as Konus-WIND. 
For a bolometric correction of $\lesssim 5$ the peak luminosity is below $10^{49}$ erg s$^{-1}$, which if related to a GRB would make EP240414a a low luminosity event. If EP240414a were a classical GRB or X-ray Flash obeying the scaling relations of \citet{amati02,yonetoku04} we would expect a spectral peak in the X-ray regime at $E_p (1+z) \sim 6$ keV and not in the $\gamma$-ray regime. 

\subsection{Light curve evolution}

\begin{figure*}[ht!]
    \centering
    \includegraphics[width=0.8\linewidth]{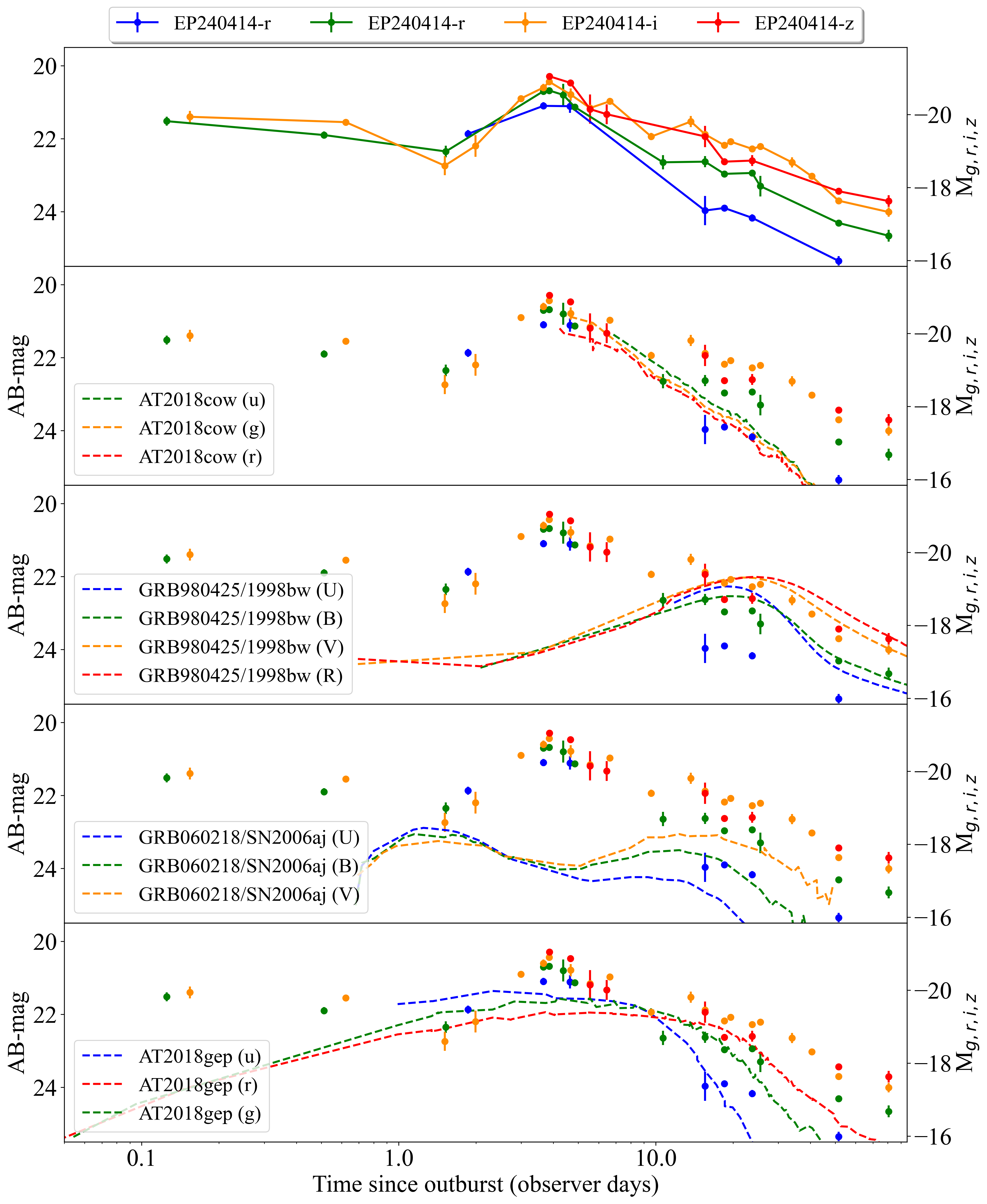}
    \caption{Light curve of EP240414a in the observer frame in \textit{g}- (blue), \textit{r}- (green), \textit{i}- (yellow) and \textit{z}-band (red) (first panel) compared to the light curves of the prototypical LFBOT AT2018cow \citep[second panel;][]{Prentice2018}, the prototypical GRB SN~Ic-BL SN~1998bw \citep[third panel;][]{2001Patat}, the X-ray flash GRB/XRF~060218 \citep[fourth panel;][]{2006Soderberg}, and the LFBOT AT2018gep \citep[fifth panel;][]{2020Ho}. From the light curve in the top panel we distinguish what we refer to as the first, second and third peak at $\sim$1, $\sim$4, and $\sim$10 days, respectively. The second panel shows the match between the steep decline in the light curve of AT2018cow and the second peak in the light curve of EP240414a. The timescale and steepness of the decline after this peak only matches AT2018cow in our selection of comparison objects. In the third panel we show that the third peak matches a prototypical GRB SN~Ic-BL like SN~1998bw well, as is similarly true for our spectroscopy at this time. However, additional components to a SN are needed to explain the first and second peak observed in EP240414a. The fourth panel compares EP240414a to X-ray flash GRB/XRF~060218 in which similarly no GRB was observed. This event shows two peaks, but at different timescales than EP2404141a. In the fifth panel we make a comparison to AT2018gep where an SN~Ic-BL appeared at late times \citep{Pritchard2021}, which shows that despite this LFBOT-SN~Ic-BL connection, its light curve does not match that of EP240414a's second and third peak well.}
    \label{fig:lccomp}
\end{figure*}

\begin{figure*}
    \centering
    \includegraphics[width=\textwidth]{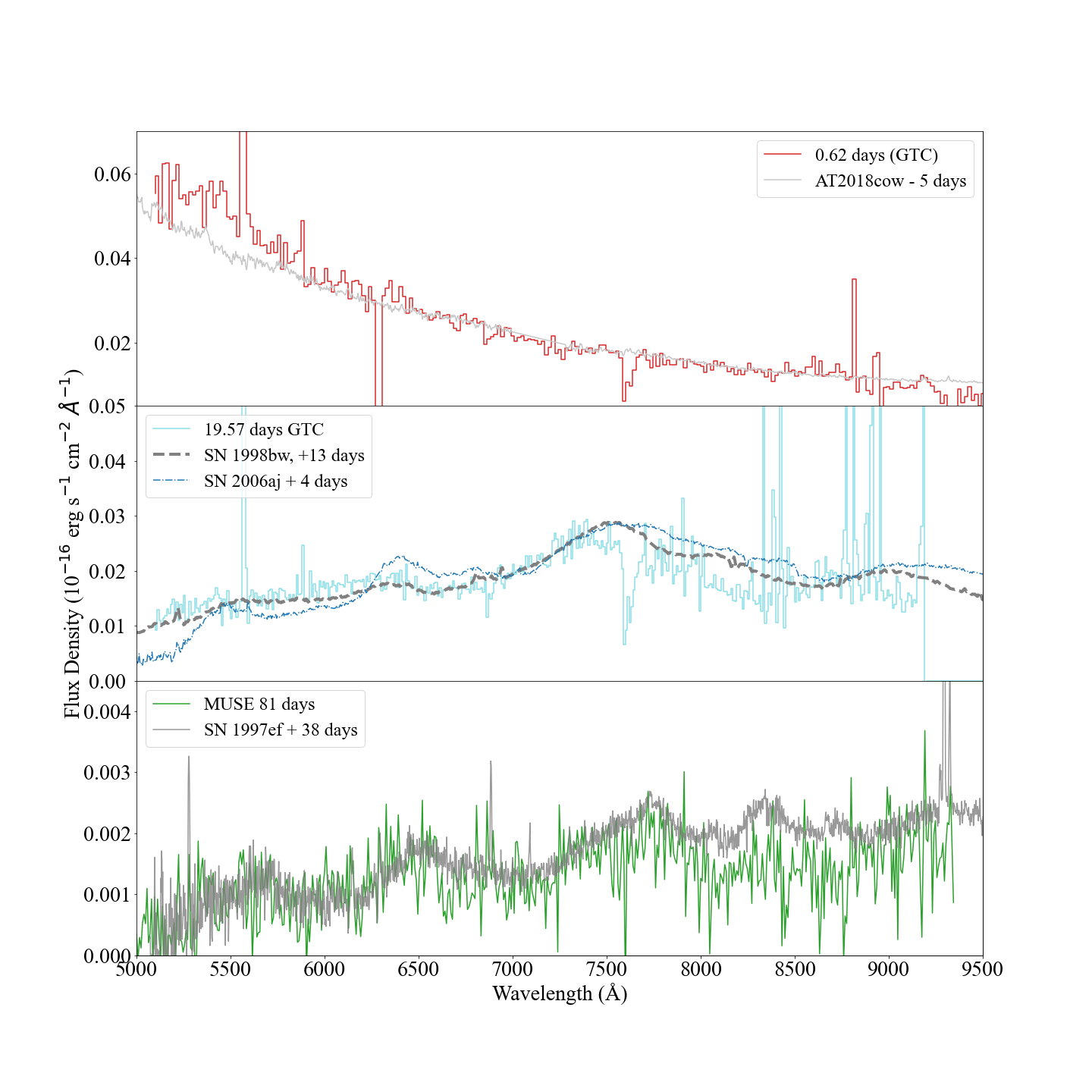}
    \caption{Three key comparisons to our spectroscopy at different epochs. The top panel shows a comparison of our 0.62~day GTC spectrum with AT2018cow at $\sim$5 days. The middle panel presents the SN~Ic-BL in the GTC spectrum taken at 19.5~day as compared to SN~1998bw at $\sim$13 days and SN~2006aj at $\sim$4 days. We obtained the best fit with SN~1998bw when fitting with both \texttt{Gelato} \citep{2008Harutyunyan} and \texttt{SNID} \citep{2007Blondin}. The bottom panel shows a MUSE spectrum extracted at the transient location at $\sim$81 days. The SN has faded significantly at this epoch. We find the best fit spectrum at this time is SN~1997ef at $\sim$38 days.}
    \label{fig:SNcomp}
\end{figure*}

The optical light curve is shown in Figure~\ref{fig:lc} and is markedly different from those of most GRBs. In particular, following the initial detection the light curve is approximately flat (albeit sparsely sampled) from 0.12 to 0.5 days post-trigger, fading by only $\Delta r = 0.38 \pm 0.16$ and $\Delta i = 0.15 \pm 0.16$ mag (for a power-law with $t^{-\alpha}$ this corresponds to $\alpha_r \sim 0.5$ and $\alpha_i \sim 0.2$). Non-detections from the subsequent night (by e.g., \citealt{Belkin2024} and this work) suggest a more rapid decline, but then the source re-brightened by about 1.3 mag between 2--3 days after the trigger, as was first reported by \citet{Srivastav2024}. This rebrightening reaches a peak absolute magnitude of $M_I = -20.9$ at 4 days before fading rapidly, its duration at half-peak brightness is $<3$ days. While rebrightening within the first day is not unheard of in GRB afterglows \citep{2018DeUgartePostigo}, the combination of brightness and timescale observed in EP240414a is unusual for GRBs. Additionally, as shown in Section~\ref{spectroscopy}, the early spectra of the source are too blue to be related to an afterglow. On the other hand, the brightness and decline from the peak of the light curve is reminiscent of the population of luminous fast blue optical transients (LFBOTs). Indeed, a comparison with the LFBOT proto-type AT2018cow (Fig.~\ref{fig:lccomp}) reveals very similar absolute magnitude and decline rates between the two events \citep{Prentice2018}, although AT2018cow is significantly bluer at the same epochs than EP240414a (see Fig. \ref{fig:SNcomp} and Section~\ref{spectroscopy}). 

The rapid decline after 4~days continues to $\sim$10~days before there is a slowing of the decline and a marginal re-brightening in the \textit{i}-band. The timescale and absolute magnitude of this re-brightening are  SNe seen in GRBs, lying somewhat between SN~1998bw \citep{1998Galama} and GRB/XRF060218/SN2006aj \citep{2006Campana, 2006Mazzali, 2006Pian, 2006Soderberg, 2006Sollerman}, an interpretation also supported by spectroscopic evidence (see Section~\ref{spectroscopy}). The optical light curve after $\sim 10$~days is therefore straightforward to explain as relating to a SN.

\subsection{Spectroscopy}\label{spectroscopy}
Spectra of EP240414a show a clear transition in the spectral shape and features as the transient evolves as shown in Figure~\ref{fig:SED}. The first spectrum by GTC $\sim$0.6~days after burst has a very blue spectral slope, if expressed as a power-law spectral index with $F_{\nu} \propto \nu^{\beta}$ then $\beta = 0.9 \pm 0.2$. In particular, we note that this is very different from the typical GRB spectral slope of $\beta \sim -1$ and too blue to be a GRB afterglow. Alternatively, the slope can be fit with a blackbody with a temperature of $T_{BB} = 24000 \pm 7000$ K. In the top panel of Figure~\ref{fig:SNcomp}, this slope is compared to the spectrum of the LFBOT AT2018cow \citep{Prentice2018}, which has a similar slope at $\sim$5 days. 

The spectral slope evolves in $\sim$3.8~days to the near featureless flat spectrum seen in the first MUSE observation shown in Figure~\ref{fig:SED}. The FORS2 spectrum taken $\sim$1~day later is similarly featureless and flat (with the exception of the bump at 4000\AA~ which is due to an imperfect flux calibration).

In observations taken more than $\sim$10~days after the burst, we observe a typical spectrum of a SN~Ic-BL. The SN features are most pronounced in the GTC spectrum taken at $\sim$19.5~days. We fit this spectrum with both \texttt{Gelato} \citep{2008Harutyunyan} and \texttt{SNID} \citep{2007Blondin} and find the best match with SN~1998bw at $\sim$13~days. In the middle panel of Figure~\ref{fig:SNcomp} we show the closest match between the GTC spectrum and SN~1998bw \citep{2001Patat} shifted to \textit{z}=0.401. 
In addition we also compare the spectrum of SN~2006aj at a phase 4 days post peak. While not formally as good a match as SN~1998bw, the overall spectral shape is unsurprisingly similar. 

Finally, we also compare the very late time (+81 day) epoch of spectroscopy obtained with MUSE. Despite the very faint source at this time, the counterpart is well detected and the broad features are clearly visible in the spectrum. A match at this epoch is best with SN~1997ef, another SN Ic BL event. We note that spectra of SN~1998bw and SN~2006aj at these epochs are not available for comparison. The presence of broad-lined Ic features in three of our spectra taken over a wide time baseline provides strong evidence of the origin of EP240414a in a massive star collapse, despite the unusual galactic location. 

\subsection{Host galaxy}\label{host}

\begin{figure*}
    \centering
    \includegraphics[width=\linewidth]{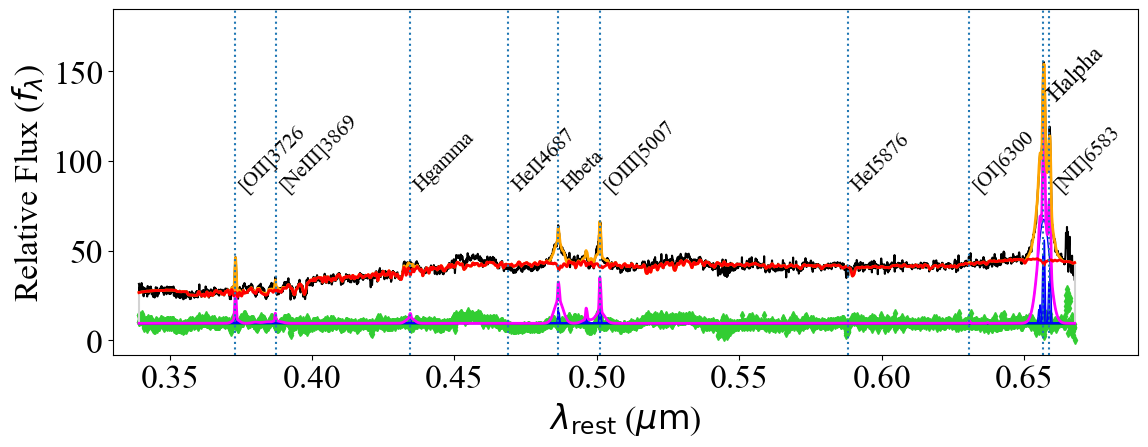}
    \includegraphics[width=\linewidth]{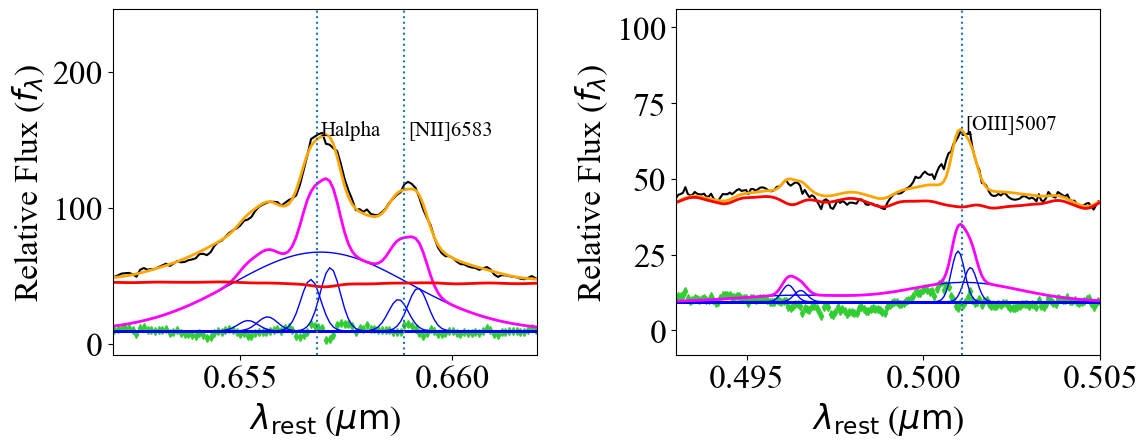}
    \caption{Best-fit \texttt{pPXF} model to the MUSE spectroscopy of the galaxy taken on 4 July; the top panel shows the full spectrum, while bottom panels show zoomed-in spectra at the locations of  the H$\alpha$ and [NII]  (Left) and (Right) [OIII]. In all cases, the observed spectrum is shown in black with the flexible stellar population synthesis template in red and several fitted emission lines in orange. Model-subtracted residuals are shown in green with the position of the emission lines on top in pink. The blue curves underneath the pink curve indicate the individual Gaussian line components.
    }
    \label{fig:muse_pPXF}
\end{figure*}

A striking feature of EP240414a is its local and wider-scale environment. The host galaxy is a luminous ($r=19.44 \pm 0.01$, $M_R = -22.5$) spiral galaxy with a large effective radius (1.5\arcsec, 8.3 kpc) at $z=$0.401. Such galaxies are rare in the long-GRB population, with none of the sample of \citet{fruchter06} and few in the low-z sample of \citet{2018Japelj} being as large, or as luminous. Indeed, amongst the long-GRB host population only the host of GRB 190829A appears similar, although it has an absolute magnitude of $M_R \sim -21.8$ and a half light radius of $\sim 6$ kpc, indicating it is substantially smaller than the host of EP240414a. Indeed, the host galaxy of EP240414a also hosts a moderately strong AGN, again unusual in the long-GRB host population \citep[e.g.,][]{2023Levan}. A higher frequency of AGNs is present in the short GRB population \citep[e.g.,][]{fong15,levan17}.

\begin{figure*}
    \centering
    \includegraphics[width=\linewidth]{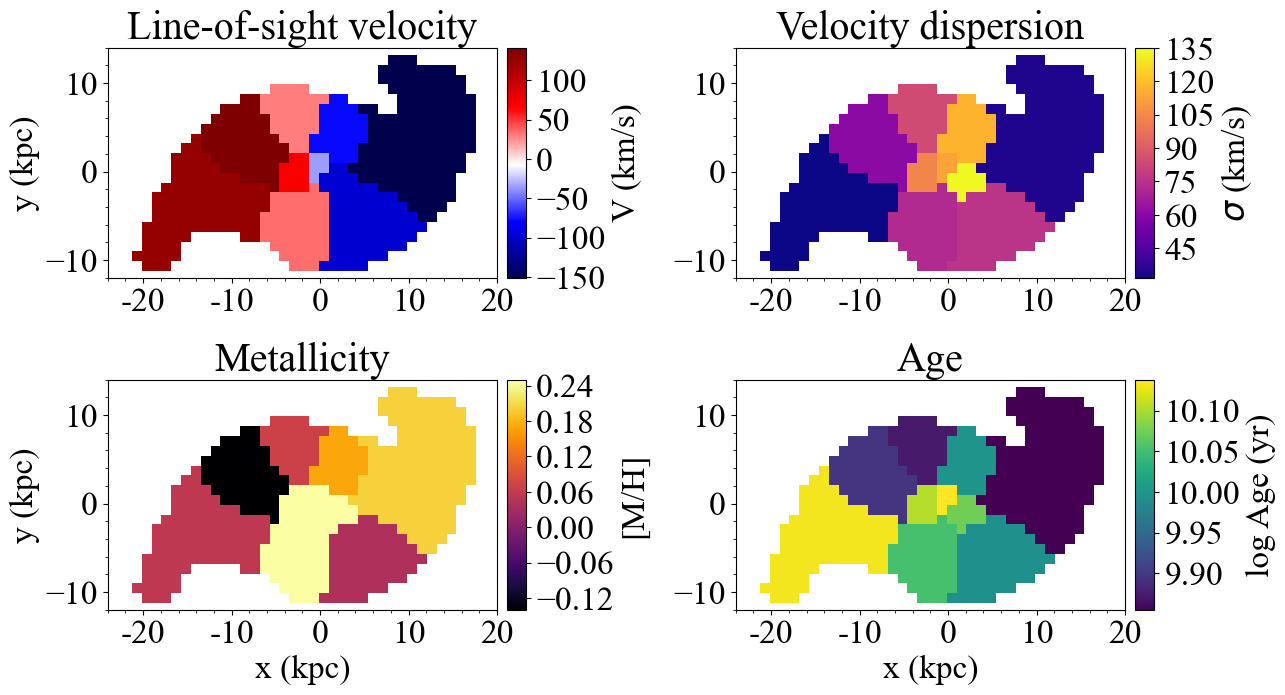}
    \caption{Spacial distributions of the velocity dispersion $\sigma$, the line-of-sight velocity \textit{V}, the metallicity [M/H] and age in the 42 Voronoi bins of the host galaxy observed with MUSE on 4 July, 2024. The axes are given in \textit{x} and \textit{y} distances from the central pixel of the galaxy in kpc.}
    \label{fig:muse_pPXF_bins}
\end{figure*}

To investigate the properties of the unusual host galaxy of EP240414a we perform detailed modelling of its spectroscopy from the MUSE data taken on 4 July. We use the penalised pixel fitting method \texttt{pPXF} \citep{2017Cappellari} to fit spectroscopy of the host which is spatially Voronoi binned to a target $S/N=10$ per bin using the \texttt{VorBin} method and software of \citet{2003Cappellari}. Each spectroscopic bin in each spaxal with a $S/N<1$ is rejected to remove residual-dominated spectra before binning. We obtain the average spectrum of the galaxy over all spatial bins and bin average spectra for each of the 42 bins covering the galaxy in the full spectral range covered by MUSE.
We use \textit{Flexible Stellar Population Synthesis} [fsps v3.2] \citep{2009ConroyI, 2010ConroyII, 2010ConroyIII} as our stellar population synthesis model template. As input parameters we use the redshift of \textit{z}=0.401$\pm$0.005 obtained from a fit to the H$\alpha$ line, and $V=0$~km s$^{-1}$ and $\sigma=200$~km s$^{-1}$ as initial guesses for the line-of-sight (radial) velocity and velocity dispersion, respectively. The fit we obtain for the average spectrum of the galaxy is displayed in the top panel of Figure~\ref{fig:muse_pPXF}. The average spectrum is shown in black with the best fit galaxy template plotted on top in red. The residuals to the fit of the galaxy template to the average spectrum with the emission masked is shown in green. As the galaxy template does not account for the presence of an AGN in the spectrum, we expect larger residuals for instance in the region between 4000\AA~and 5400\AA~if there is broad Fe~II emission \citep{1992Boroson}. We observe these larger residuals which hint at broad features from the AGN and encourage more detailed modelling of the AGN emission which goes beyond this work. The emission lines in the average galaxy spectrum are fitted with Gaussians together with the galaxy template. We account for all emission lines included in \texttt{pPXF} and find  H$\alpha$, H$\beta$, H$\gamma$, [Ne~III]~(3869~\AA) and the doublets [O~II]~(3729~\AA), [O~III]~($5007$~\AA), and [N~II]~($6583$~\AA) in our model of the galaxy emission lines. Emission lines are shown in orange and their rest wavelength is marked with dashed blue lines. We do not detect He~II~($4686$~\AA), He~I~($5876$~\AA), and the [O~I]~($6300$~\AA) doublet in the host galaxy spectrum. In the bottom panel of Figure~\ref{fig:muse_pPXF} we show the best fit to the $H\alpha$ and the [N~II]~($6583$~\AA) doublet and [O~III]~($5007$~\AA) lines. From the ratios of the flux [O~III]/$H\beta$ and [N~II]/$H\alpha$ we determine the location of the galaxy in the Baldwin–Phillips–Terlevich (BPT) diagram. It falls in the composite galaxy part of the diagram  \citep{1987Veilleux, 2013Vitale}. The emission-line spectrum originates from both star formation and the AGN as we are taking into account the average spectrum of the whole galaxy. 

We estimate the total mass of the galaxy from the average velocity dispersion for which we find $\bar{\sigma}=119\pm3$~km s$^{-1}$ from our fit. Using the relation $\sigma=\sqrt{\frac{GM}{CR}}$ with C=2.25 for rotation dominated dispersion and R the of the galaxy \citep{2009Epinat}, we find $M=(7.0\pm0.3)\times10^{12}M_{\odot}$. The mass to light ratio we obtain from our modelling is 2.6 for the $r$-band light which gives a stellar mass of $M_{*}=(2.21\pm0.01)\times10^{11}M_{\odot}$ for an $r$-band luminosity of $L=(8.55\pm0.04)\times10^{10}L_{\odot}$ using the $r$-band magnitude from Table~\ref{tab:host}. We calculate the color excess E(B-V) from the Balmer optical depth by measuring the ratio of $H\alpha$ to $H\beta$ and using a Balmer decrement of 3.1 which is often assumed for AGN \citep{osterbrock2006astrophysics}. We find E(B-V)=0.54 from the average spectrum. We clearly observe broad features in the spectrum with for example a velocity of $\sim$1600~km s$^{-1}$ for the H$\alpha$ line which indicates this is a Syfert 1 galaxy. The color excess is notably higher than in typical Syfert 1 galaxies although not unprecedentedly high (\citealt{2004Hopkins}). 

When considering spacially resolved, individual, bins the central bin shows evidence for broad lines likely from the AGN, while the spiral arms show narrow features associated with star formation. From the narrow component of the $H\alpha$ line in the bins covering the spiral arms, we can estimate the star formation rate (SFR) of those parts of the galaxy using the \citet{1998Kennicutt} relation assuming a Salpeter IMF. We find a SFR of 2.15 $M_{\odot}$/yr and note that in excluding the central region where the AGN dominates the light might lead us to miss a portion of the star formation close to the centre making this a lower limit on the SFR. 

In Figure~\ref{fig:muse_pPXF_bins} we show the spatial distributions for $\sigma$, \textit{V}, the total metallicity [M/H], and age. \textit{V} in the top left panel reveals the structure of a spiral galaxy.
From the top right panel of the velocity dispersion, we see that the $\bar{\sigma}$ of $\approx 120$~km s$^{-1}$ is determined by the $\sigma$ of the central part of the galaxy which is brightest and thus the dominant component in the average spectrum. The higher velocity dispersion and higher metallicity in the centre are due to the presence of the AGN. In the bottom left panel we find that there are a few low metallicity regions in the spiral arms of the galaxy. There are only small differences in the stellar age of the different parts of the galaxy as shown in the bottom right panel.

\begin{figure}[ht!]
    \centering
    \includegraphics[width=\linewidth]{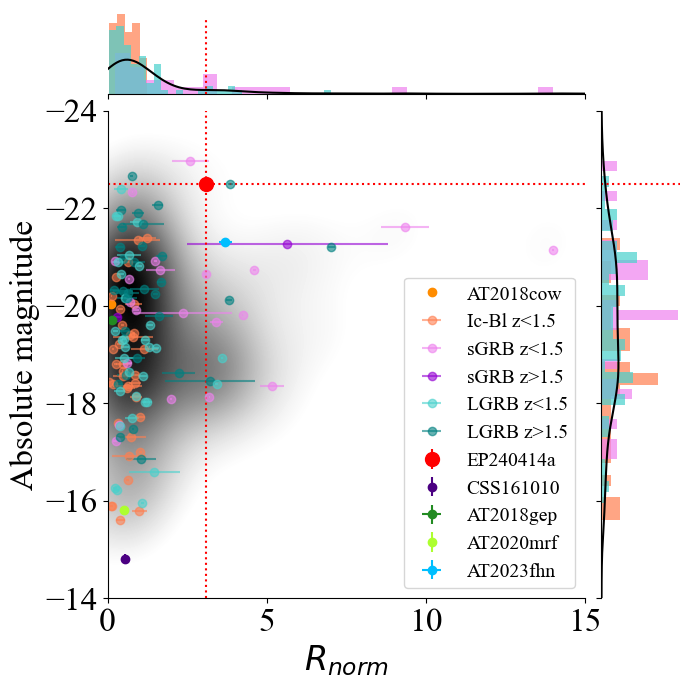}
    \caption{Absolute magnitude versus host normalised offset $R_{norm}$ for long-GRBs \citep{2002Bloom, 2016Blanchard}, short-GRBs \citep{2022Fong}, SN Ic-BL with and without GRB \citep{2018Japelj}, and LFBOTs with reported host-offsets (AT2018cow \citet{Prentice2018}; CSS161010 \citet{2020Coppejans}; ZTF18abvkwla \citet{2020Ho}; AT2020mrf \citet{2022Yao}; and AT2023fhn \citet{2024Chrimes}) and EP240414a. At the top and right hand side of the plot the distributions in $R_{norm}$ and absolute magnitude of the GRBs and the SNe~Ic-BL are shown, respectively. The black solid curves on top of these distributions indicates the density of the sum of these distributions similar to the 2D density underlying the populations in the main figure. When compared to the long-GRB population and the SN~Ic-BL population, EP240414a has a large normalised offset, which is more in line with that of short-GRBs. In the small sample of LFBOTs, AT2023fhn has a normalised offset similar to that of EP240414a. The absolute magnitude of the host galaxy of EP240414a is similar to that of the brightest hosts of GRBs and LFBOTs.}
    \label{fig:hostoffset}
\end{figure}

The local environment of the burst is similarly noteworthy. As mentioned above, the projected physical offset is 25.7 kpc. The host normalised offset of $R_{norm} = r_{transient}/r_e$ = 3.1 is larger than seen in any other SN GRB to date (here $r_{transient}$ is the offset of the transient from the centre of light of the host and $r_e$ the effective half-light radius of the host galaxy). In Figure~\ref{fig:hostoffset} we show the absolute magnitude versus the host normalised offset for the host galaxy of EP240414a compared to that of long-GRBs, short-GRBs, SN~Ic-BL, and LFBOTs. The host normalised offset is considerably larger than that of long-GRB and SN~Ic-BL and more similar to that of short-GRB hosts. The long-GRBs with larger host normalised offsets, notably GRB 230307A with a 40 kpc offset \citep{levan2024b}, these have been generally associated with compact object mergers rather than collapsar GRBs. Nevertheless, the detection of a SN~Ic-BL in EP240414a clearly indicates the collapse of a massive star as the progenitor. 

\begin{figure}[ht!]
    \centering
    \includegraphics[width=1.1\linewidth]{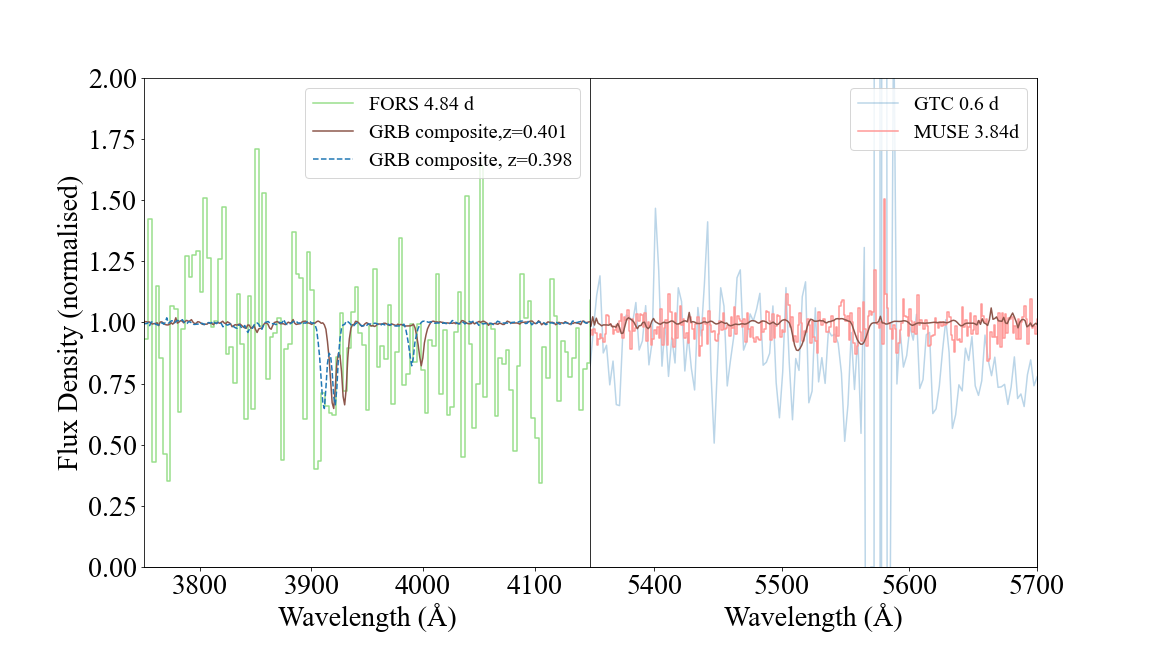}
    \caption{Zoom in of our spectra around the expected locations of strong absorption lines, in particular the MgII doublet in our blue FORS2 spectrum and Ca H and K in our MUSE observations. In addition to our observations we also overplot the composite GRB afterglow spectrum from \citet{christensen} as a comparison. We note that MgII at $z=0.401$ does not match, an offset to $z=0.398$ (500 km s$^{-1}$) would line up with some strong features in the FORS2 spectrum. The low signal-to-noise ratio in the spectrum means that the lack of strong absorption does not imply an unusually low density line of site. For Ca H \& K the observations are also not sufficiently sensitive to strongly constrain the absorption properties, although they do imply that any absorption is likely somewhat less pronounced than in the average GRB spectrum.  }
    \label{fig:abs_lines}
\end{figure}

The late-time spectroscopy obtained with MUSE provides tentative evidence for the detection of an H$\alpha$ emission line at the location of the transient, suggesting weak underlying star formation. There is no velocity offset between this line and the emission line from the spiral arm that extends in the direction of the transient (but does not visibly underly it). However, given the point spread function, this could be related to light from the spiral arm which leaks into the aperture at the source location.

In principle, the early afterglow spectroscopy should also enable the line of sight to the transient to be studied in absorption. However,  there are no prominent absorption features detected in the early spectra. In particular, we do not detect Ca H\&K or Na D in absorption in our early GTC or MUSE spectra (see Figure~\ref{fig:abs_lines}), although some stronger absorption lines which might be detectable with our signal-to-noise would lie blueward outside the range covered by our OSIRIS+ and MUSE spectra. 
We do detect a plausible low signal-to-noise ratio absorption from Mg II in the FORS2 spectrum at 4-days, although note this would require $z=0.398$ for the absorbing gas for instance due to the gas moving in our direction by about 500~km s$^{-1}$ (see also Figure~\ref{fig:abs_lines}). If real, the equivalent width of these lines is $8.1\pm2.5$~\AA\ which gives $5.8\pm1.8$~\AA\ in the rest frame. A comparison can be made to the typical value of the equivalent width measured from GRB afterglow spectra of $3.76\pm0.30$~\AA\ and their line strength as described in \citet{2012DeUgartePostigo}. We find that with a line strength parameter of 0.63, the Mg~II feature in our spectrum is stronger than ~80\% of the features in GRB spectra in their sample.  Therefore, the lack of significant absorption lines may arise from the low signal-to-noise spectroscopy and the lack of access to the stronger lines that lie in the UV at the transient redshift. It does not necessarily imply a very unusual (e.g., low-density) environment.

\section{Discussion}\label{discussion}

We obtained extensive follow-up optical and near-infrared photometric and spectroscopic observations of EP240414a in the days to weeks after the event, revealing a rapidly evolving transient at $z=0.401$, which is offset 25.7 kpc from its host galaxy. At late times ($>10$ days) clear signatures of an SN~Ic-BL appear which are an excellent match to those seen in collapsar-driven GRBs. 

Our observations lead to the simplest explanation of EP240414a, which is that it is a low luminosity collapsar GRB observed at a higher redshift than typical for such bursts because of the sensitivity of the EP-FXT. At this low luminosity, an X-ray-rich burst is expected (or, more formally, an X-ray Flash), and the combination of expected softness and faintness do not cause tension with the lack of a reported $\gamma$-ray signal. 

Whilst the prompt high-energy and late-time optical/IR emission can be remedied straightforwardly within this model, the observed behaviour on timescales of a few days is very different from that typically seen in GRBs, even those of low luminosity. The luminous bump at $\sim 4$ days has similar energetics and temporal evolution to the LFBOT population, although is not as blue. Hence, developing a complete picture of the entire emission from EP240414a is not as straightforward. Below we first consider the overall light curve and spectral evolution in context with other events and then discuss plausible physical explanations for the observed properties. 

\subsection{EP240414a in context}
\label{sec:context}

To better understand how the multi-wavelength counterpart of EP240414a connects with the zoo of optical/IR transients we make several comparisons with those populations, paying particular attention to the SNe associated with GRBs (e.g., GRB 980425/SN~1998bw \citealt{1998Galama}, GRB/XRF 060218/SN~2006aj \citealt{Campana2006a,2006Pian}) and the LFBOT population (e.g., AT2018cow \citealt{perley18,Prentice2018}, AT2018gep \citealt{Ho19}) as the closest matches to our photometry and spectroscopy.

\begin{figure}[ht!]
    \centering
    \includegraphics[width=1\columnwidth]{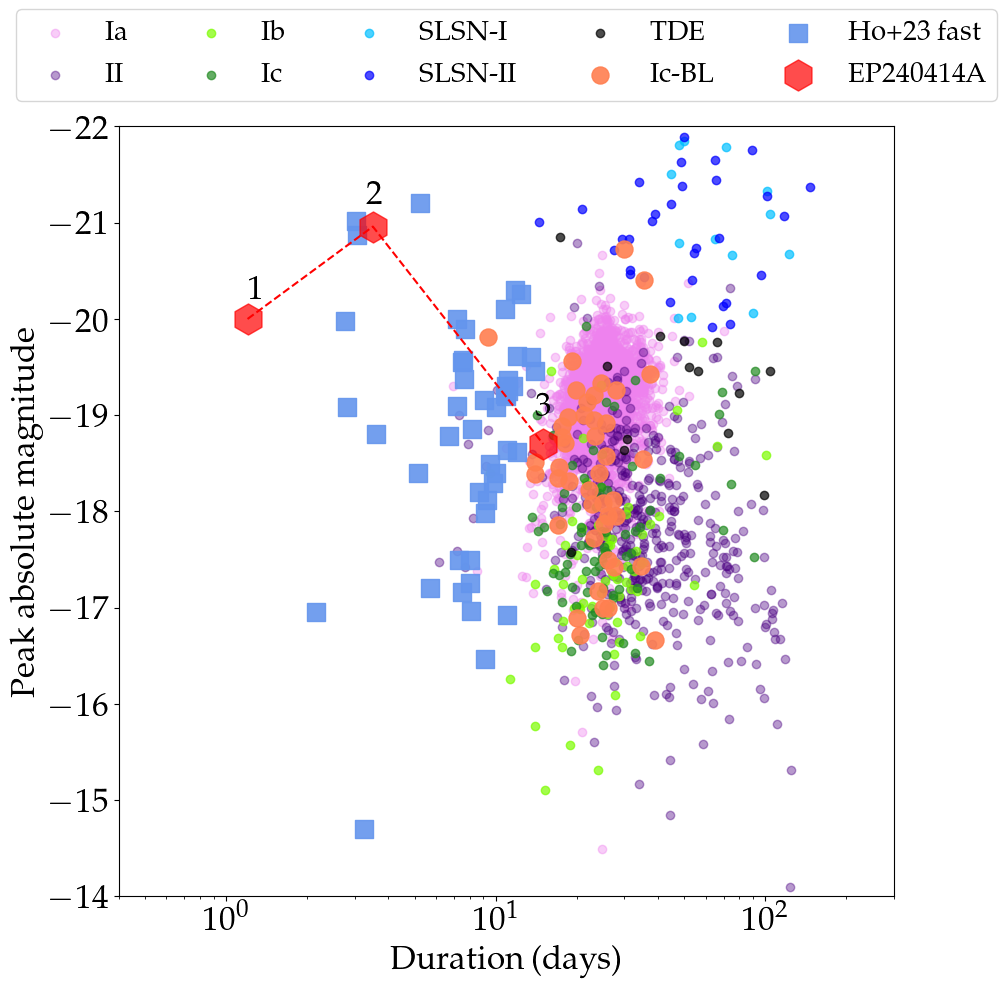}
    \caption{ Absolute magnitude at peak versus the duration of transients above half-peak magnitude updated from \cite{perley20}. For the background we show different SNe and TDEs from the ZTF bright transient sample, as well as the set of fast transients identified by \cite{Ho2023}. We note there is a clear observational selection against objects that lie towards the short ($\lesssim$~2~days) duration region because of the cadences of current surveys. We mark the approximate locations of the optical/IR light of EP240414a for each of the three peaks we identify from the light curve (where the duration and absolute magnitude of the first peak is most uncertain), demonstrating how it compares to the most luminous and fast optical transients observed to date. The implication is that some fast optical transients may arise from similar events to EP240414a, but the initial X-ray outburst was missed.}
    \label{fig:dur_peak}
\end{figure}

A common representation of the properties of a given transient is its location in the duration -- peak luminosity (or absolute magnitude) plane \citep[e.g.,][]{kasliwal11}. Placing EP240414a within this regime is non-trivial because it contains at least three separate emission episodes, although approximations can be made, and the three approximate locations are shown in Figure~\ref{fig:dur_peak}, demonstrating the rapid and luminous nature of EP240414a. The first peak is poorly sampled but likely takes place within the first few hours of the event. The earliest reported optical observations from \cite{Aryan2024} report a source with $r=21.52 \pm 0.12$ and $i=21.40 \pm 0.16$ in observations at 0.12 and 0.15 days, respectively. However, the observations do not distinguish if the source was brightening or fading during this time. Observations at $\sim 0.5$ days suggest modest fading from this point, but they are too sparse to map the morphology of the light curve in detail over this time (e.g., gradual decline, brightening and then fading). By 1.5 days there is more substantial fading observed. Hence, we approximate the peak as $M_i \sim -19.8$ at 0.13 days with a timescale of $<1$ day, although it is plausible that the peak could be brighter and the timescale shorter. The second is better mapped out by observations, with a second peak at $\sim 4$ days at $M_r \sim -20.7$ and a duration of $\sim 4$ days. The third component (or peak) is not entirely obvious in the light curve which is likely the combination of a rapidly fading component and a rising SN, which we estimate to be at around 15 days with an absolute magnitude of $M_r \sim -18.7$ and a duration of $\sim 15$~days.

The third peak in the light curve of EP240414a is most readily explained by a SNe~Ic~BL, similar to the type of SN seen in collapsar long-GRBs. It has a peak absolute magnitude and timescale that are consistent with an average rise time of $\tau_{rise}$ = 14.0 $\pm$ 0.8 days and absolute magnitude of M$_{r}$ =  -18.51 $\pm$ 0.15 mag according the sample of SNe~Ic~BL by \citet{2024Srinivasaragavan}. It appears to be somewhat more rapidly evolving than the proto-typical SN~Ic-BL associated with a GRB SN~1998bw, but similar in decline rate to SN~2006aj, seen in coincidence with the very long duration XRF~060218, also a low luminosity event. The SN in EP240414a is more luminous than SN~2006aj by a factor of $>2$. 

To compare GRB/XRF~060218/SN~2006aj to EP240414a as XRFs without GRBs, we calculate what the peak X-ray flux of XRF~060218 would be if it had occurred at a redshift of $z=0.401$ instead of $z=0.033$. It would have been $\sim4.6\times10^{-11}$~erg~s$^{-1}$~cm$^{-2}$, i.e., $\sim$250 times fainter than at $z=0.033$. This flux would just be detectable for EP, which can reach a $5\sigma$ detection of a source with a flux of $2.6 \times 10^{-11}$ erg s$^{-1}$ cm$^{-2}$ within 1000~s in the 0.5-4~keV band, although it would have been much fainter than EP240414a. It should additionally be noted that the peak flux for XRF~060218 was recorded in the 0.3–10 keV band and that the event peaked at 4.9~keV which is outside the range of EP. In the case of EP240414a, the presence of luminous X-rays (as the trigger) and the radio afterglow \citep{Bright2024} implies that this progenitor is more akin to those seen in GRBs.

\begin{figure}[ht!]
    \centering
    \includegraphics[width=1.0\linewidth]{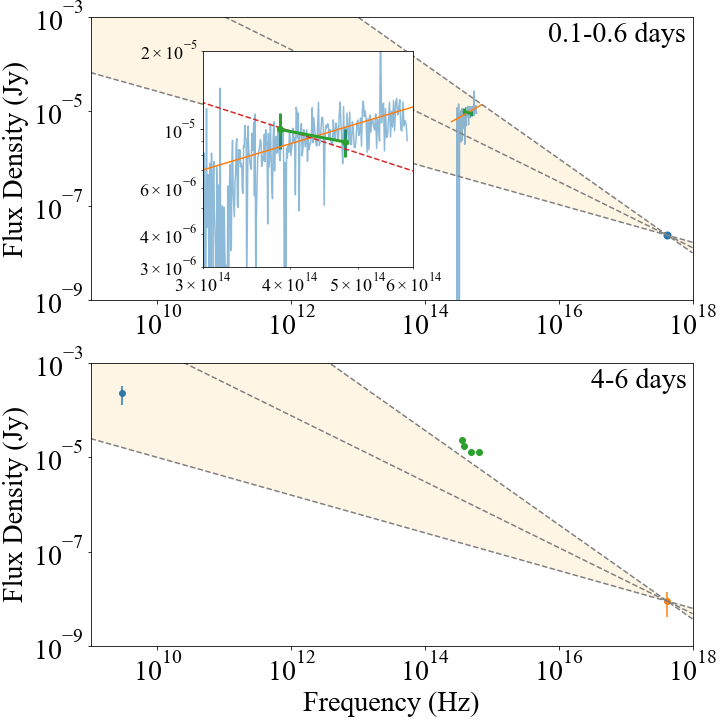}
    \caption{The evolving X-ray to radio spectral energy distribution of EP240414a seen at $<1$ day (during the first optical peak) and at 4-6 days (during the second, brightest optical peak). There are several crucial insights. Firstly, although the optical at early times is potentially consistent with the extrapolation of the X-ray spectrum shown by the yellow region, which could indicate an afterglow origin at 0.1 days, by 0.6 days the optical spectrum is inverted and cannot lie on the extension of the X-ray light, indicating we are not observing afterglow at this epoch. At 4-6 days the X-ray to optical spectral index has changed such that the optical/IR is not consistent with the expectations of the X-ray flux at this epoch. This indicates that the brightest peak is also not related to the afterglow. However, at this epoch a relatively bright radio transient was observed \citep{Bright2024}, which implies that while the optical light is a separate component there is a substantial non-thermal (jet) afterglow in addition at other wavelengths. }
    \label{fig:bbsed}
\end{figure}

\begin{figure*}
    \centering
    \includegraphics[width=0.5\linewidth]{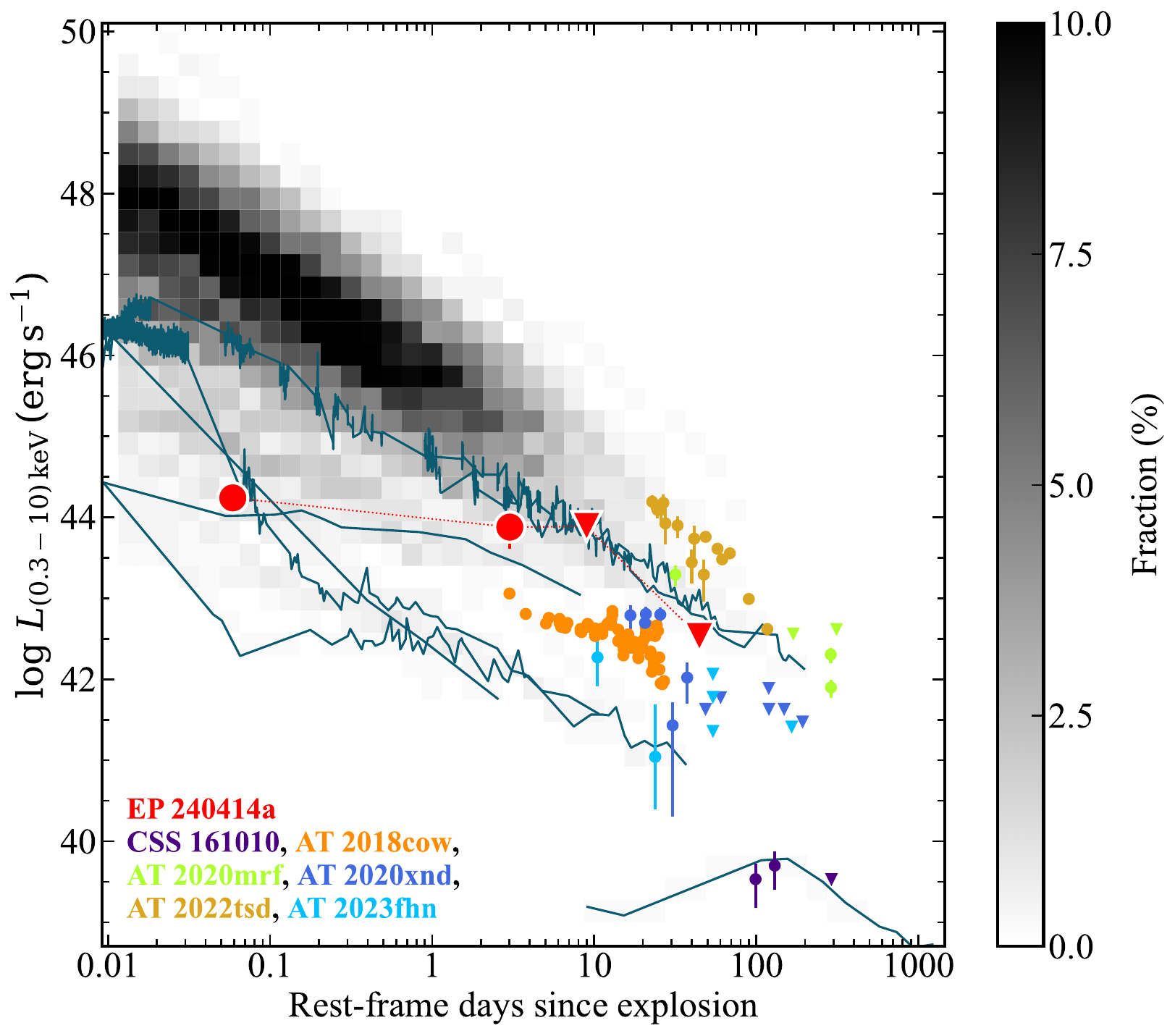}%
    \includegraphics[width=0.45\linewidth]{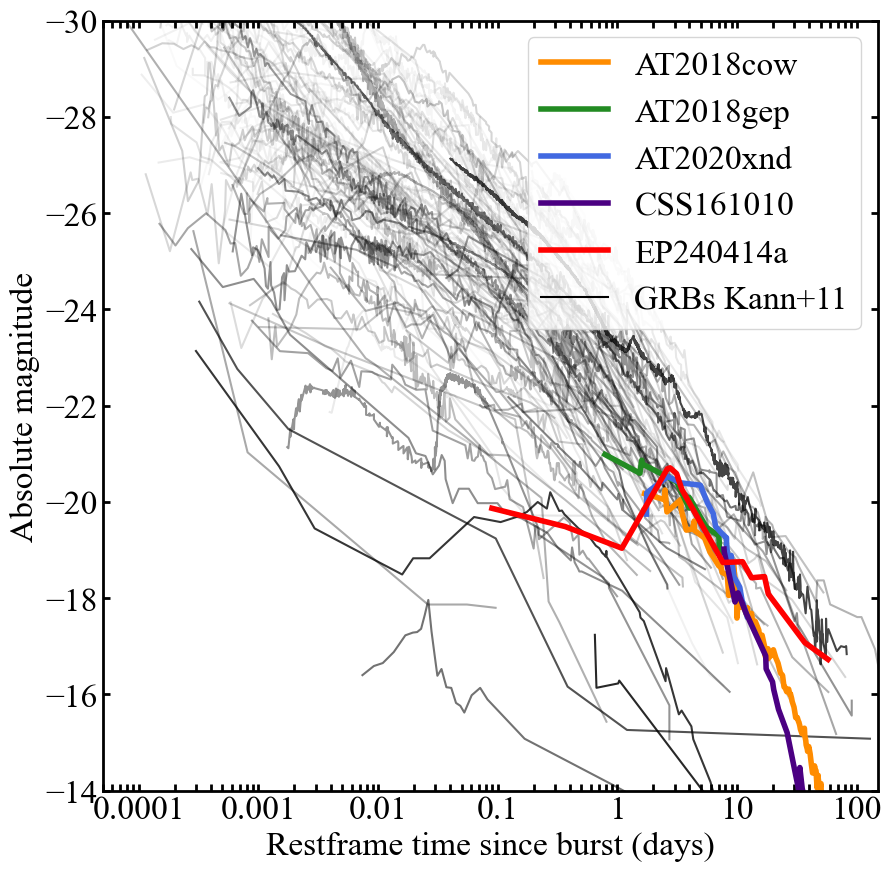}
      \caption{\textit{left:} The X-ray light curve of EP240414a (red) compared to GRBs and LFBOTs. The first measurements which we include was reported \citet{Guan2024}. The later points are the \textit{Swift} and \textit{Chandra} observations described in this work. The AGN contribution was subtracted from these measurements using the luminosity measured by \textit{Chandra}. For comparison we show the X-ray light curves of LFBOTs as adapted from \citet{2024Chrimes2} using data from \citet{2018Rivera, 2020Coppejans, 2022Bright, 2022Yao, 2023Matthews, 2024Chrimes}. The density plot shows the parameter space occupied by the GRBs; highlighed are the light curves of the low-luminosity GRBs GRB060218 \citep{Campana2006a}, 100316D \citep{Starling2011a}, and 171205A \citep{Izzo2019a}, the intermediate luminosity GRBs 120422A \citep{Schulze2014a}, 130702A \citep{Singer2013a} and 190829A \citep{Dichiara2022a}, and the off-axis binary neutron star merger driven short GRB 170817A \citep{Hajela2022a}.
      \textit{Right:} \textit{r}-band optical light curves of EP240414a (red) compared to GRBs and LFBOTs AT2018cow \citep{2021Xiang}, CSS161010 \citep{2024Gutierrez}, AT2018gep \citep{2020Ho}, AT2020xnd \citep{2021Perley}. The GRB light curves in black are obtained from \citet{kann06,kann10,Kann2011, 2012Nicuesa}.}
    \label{fig:LFBOT}
\end{figure*}

Early observations of SN~2006aj also showed an early, sub-day peak, thought to be related to shock breakout, perhaps through a dense wind. The first peak of emission for EP240414a is again somewhat more luminous (a factor $\sim 2$), but could plausibly be related to a similar event. Indeed, the rising spectrum to the blue at this epoch is consistent with the spectrum seen in LFBOTs, although at a shorter timescale (see Figure~\ref{fig:SED}). An alternative would be that the early emission represents a GRB-like optical afterglow. However, the color of the counterpart as seen in our spectroscopy at 0.6 days is inconsistent with an afterglow origin (see Figure~\ref{fig:SED} and Figure~\ref{fig:bbsed}). 

Finally, the second peak at $\sim 4$~days is perhaps the most challenging to explain. While some GRBs do exhibit outbursts on this timescale, these are often ascribed to refreshed shocks which re-energize the blastwave across a wide range of frequencies. However, while the optical light brightens by a factor of several, the sparsely sampled X-ray light curve indicates a decline from 0.1 to 4 days (see Figure~\ref{fig:LFBOT}). Indeed, the X-ray to optical spectral index changes from $\beta_{OX} = 0.85 \pm 0.05$ at 0.1 days to $\beta_{OX} = 1.1 \pm 0.1$ at 4 days, again indicating (albeit at modest significance) that an afterglow origin is unlikely for this peak (see Figure~\ref{fig:bbsed}).

Both the timescale and the luminosity of this second peak are broadly consistent with those of AT2018cow and the overall LFBOT population. However, the colors at this epoch are not as extreme as for LFBOTs, indicating a lower temperature and indeed a non-thermal spectrum, given the apparent flatness of the spectra obtained around peak. Although EP240414a is not a perfect match, this possible link to the LFBOT population is intriguing, since it implies that similar progenitors could be responsible for LFBOTs and FXTs. While the link to short-lived high-energy (GRB-like) emission is novel, there has been one event which shows a rapid rise to LFBOT-like luminosity, and a subsequent broad-lined type Ic supernova. This object, AT2018gep \cite{Ho19}, is also plotted in Figure~\ref{fig:lccomp}. In comparison to AT2018gep, the counterpart of EP240414a is somewhat brighter and faster evolving, with a clearer separation of the $\sim 4$ day peak and later, presumably nickel driven, supernova emission.

The comparison between EP240414a, GRBs, and LFBOTs at X-ray and optical wavelengths is made in Figure~\ref{fig:LFBOT}. The left panel shows the X-ray light curves of all \textit{Swift} GRBs detected up until mid-February 2024 as a density plot, which is made with data retrieved from the Swift Burst Analyser \citep{Evans2010a}. We selected the GRBs with at least 2 detections with \textit{Swift}/XRT and a known redshift. The total sample consists of 484 long and short GRBs. We processed their light curve data and moved them to their rest-frames following \citet{Schulze2014a}. The LFBOT light curves in the left panel are adapted from \citep{2024Chrimes2}. The X-ray light curve of EP240414a is not clearly distinct from those of LFBOTs within a 3-300 day window, although strikingly the LFBOT population (especially given the uncertainty in explosion time in most cases) are also not clearly distinct from the late time X-ray light curves of GRBs. 

\begin{figure}
    \centering
    \includegraphics[width=1.1\linewidth]{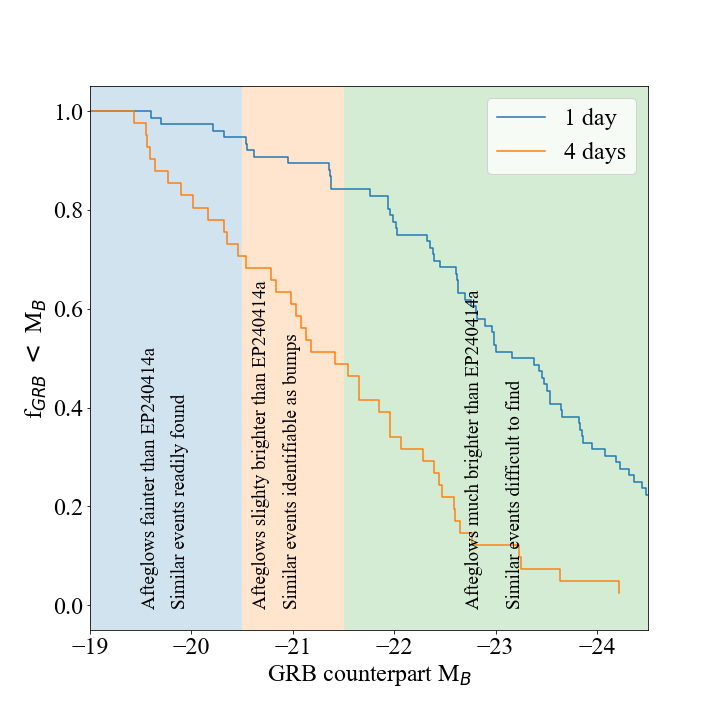}
    \caption{Absolute magnitude of GRB afterglows at 1 and 4 days post-burst, compared with the magnitudes of EP240414a at the same epochs. The implication is that emission like the first episode (which has spectra similar to LFBOTs) could be readily hidden, but the 4-day peak would only be straightforward to hide in a minority of bursts.  }
    \label{fig:brightness}
\end{figure}

One  question raised by the identification of LFBOT-like emissions in EP240414a is if such components could be common in more energetic GRB afterglows, but missed because they are lost in the glare of the afterglow light. In Figure~\ref{fig:brightness} we show the distribution of absolute magnitude of GRB afterglows at 1 and 4 days after the GRB detection taken from \cite{Kann2011}. At one day almost all GRBs were brighter than $\sim -19.5$ and so it is likely that emission similar to that seen in the early epochs of EP240414a (and GRB/XRF 060218) could be a frequent component in GRBs, but is missed because of bright afterglows. However, at the time of the second emission episode at $\sim 4$ days, only 50\% of GRB afterglows are brighter than $M \sim -21$, and so we would expect to have observed LFBOT-like emission should it arise. 

\subsection{Physical origin of EP240414a}

Our observations of EP240414a suggest a causal link between long-GRBs, FXTs and some LFBOTs. The presence of SN emission consistent with SN~Ic-BLe such as SN~1998bw post $\sim 20$ days demonstrate that the progenitor of EP240414a was a massive star undergoing core-collapse. That this SN has similar properties to SNe accompanying other GRBs implies that the star itself is unlikely to be very different from those which create the bulk of the long-GRB population. The difference in the properties of the counterpart that we observe is therefore most likely related to factors extrinsic to the star itself, such as our viewing angle, or the immediate environment in which the star lies.

The connection to a collapsar long-GRB immediately provides a few plausible physical mechanisms for the various observations of EP240414a. Long-GRBs involve strongly relativistic outflows that are beamed into a small fraction of the sky. Classical, highly energetic long-GRBs are then viewed close to the jet-axis. Depending on the structure of the jet an observer at different off-axis angles could see a very different event, from no-GRB \citep[in the case of a top-hat jet structure; e.g.,][]{Rhoads1999} to a much weaker and potentially softer GRB \citep[for highly structured jets; e.g.,][]{lamb17}. In principle, this may explain both the prompt X-ray emission with the lack of a $\gamma$-ray detection, with the radio detection and optical light curve (before 20 days) as product of the forward shock produced due to the relativistic jet interacting with the circumstellar-medium (CSM) around the progenitor star~\citep{Sari1998}. In practice, the multi-peak nature of the optical light curve and spectrum is inconsistent with expectations for a forward shock (see Fig.~\ref{fig:bbsed} and discussion above), even accounting for off-axis observers. Given the multiple peaks and general temporal evolution, a combination of a reverse and forward shock provides a more plausible alternative. However, again, the spectrum at 0.6 days is in tension with this expectation. So, while the radio observations indicate synchrotron emission from a relativistic jet \citep{Bright0414a}, this is likely not dominant in the optical light curve, perhaps at any epoch.

It is also relevant to consider if the location of the event within the host galaxy could have any bearing on the progenitor or its physics. In particular, the large offset from the host galaxy is uncommon in long GRBs as can be seen in Figure~\ref{fig:hostoffset}. An underlying, faint star-forming region could explain this, but would again be at an unusual location within its host galaxy. We observe H$\alpha$ emission close in projection to the transient location without a velocity offset, however it is unclear if this emission is associated with a star formation region, or something else. If there is no star formation directly at the location of the event, then it is extremely difficult to explain how an event typically associated with massive stars could have occurred at such an remote location. In particular, obtaining a distance of tens of kpc from the nearest star formation will take $t_{10 kpc} =$ 100 Myr $\times v/100$ km s$^{-1}$ (for ballistic travel, ignoring the likely substantial effect of the galactic potential). The long lifetimes of more than 100 Myrs are unlikely to be reached for the massive stars which could form GRBs. Even for binaries which are kicked from their birth site during the first SN, and then create a GRB during the second SN a lifetime of tens of Myrs would be substantial. However, it should also be noted that detailed population synthesis calculations do include SN-like events at such large evolution times, albeit in the minority \citep{2017eldrige}. The challenge here would be in explaining why such an unusual channel would produce a SN so similar to SN~1998bw. 

The link to a massive star progenitor and the presence of relativistic jet emission in the optical regime (it is likely the X-rays and radio do arise from at least a moderately relativistic jet) suggests another possibility for the origin of the optical counterpart of EP240414a; namely, interaction of the jet and SN ejecta with the envelope of the progenitor star. This interaction creates a combination of broad shocked stellar material (observable at large viewing angles) and narrower shocked jet material. This so-called cocoon, produces a wide range of EM signatures~\citep{Nakar2017, Piro2018}. In particular, the shocked-jet material could produce the early X-rays seen in EP240414a without a $\gamma$-ray detection, with the shocked stellar material responsible for the early optical light curve, and explaining the very blue early spectrum. This may provide a viable explanation for the early optical emission ($<$~1~day), which is substantially fainter than long-GRB afterglows at the same epoch and could readily be missed in many GRBs but observable for off-axis observers which have weak afterglows at this epoch (see Figures~\ref{fig:LFBOT} and \ref{fig:brightness}). Detailed predictions for cocoon emission depend sensitively on the degree of mixing between the jet and the cocoon \citep{Nakar2017}, but are broadly consistent with the observations of the first peak. The cocoon emission provides a less natural explanation for the second, most luminous peak at $\sim 4$ days because it implies that such cocoon interactions must be rare in GRBs to have evaded detection so far, since observations would be sensitive to such bumps in $\sim 50\%$ of GRBs (see Section~\ref{sec:context}). However, given that SNe~Ic~BL can be born both with and without relativistic jets, it could also be that EP240414a represents a bridge between the two scenarios in which the jet properties differ from those in normal GRBs. Regardless, it is difficult to reconcile the second-peak timescale with expectations of cocoon emission, which is expected to peak at $1$ day for typical parameters~\citep{Nakar2017, Piro2018}.

\begin{figure*}[ht!]
    \centering
    \includegraphics[width=\linewidth]{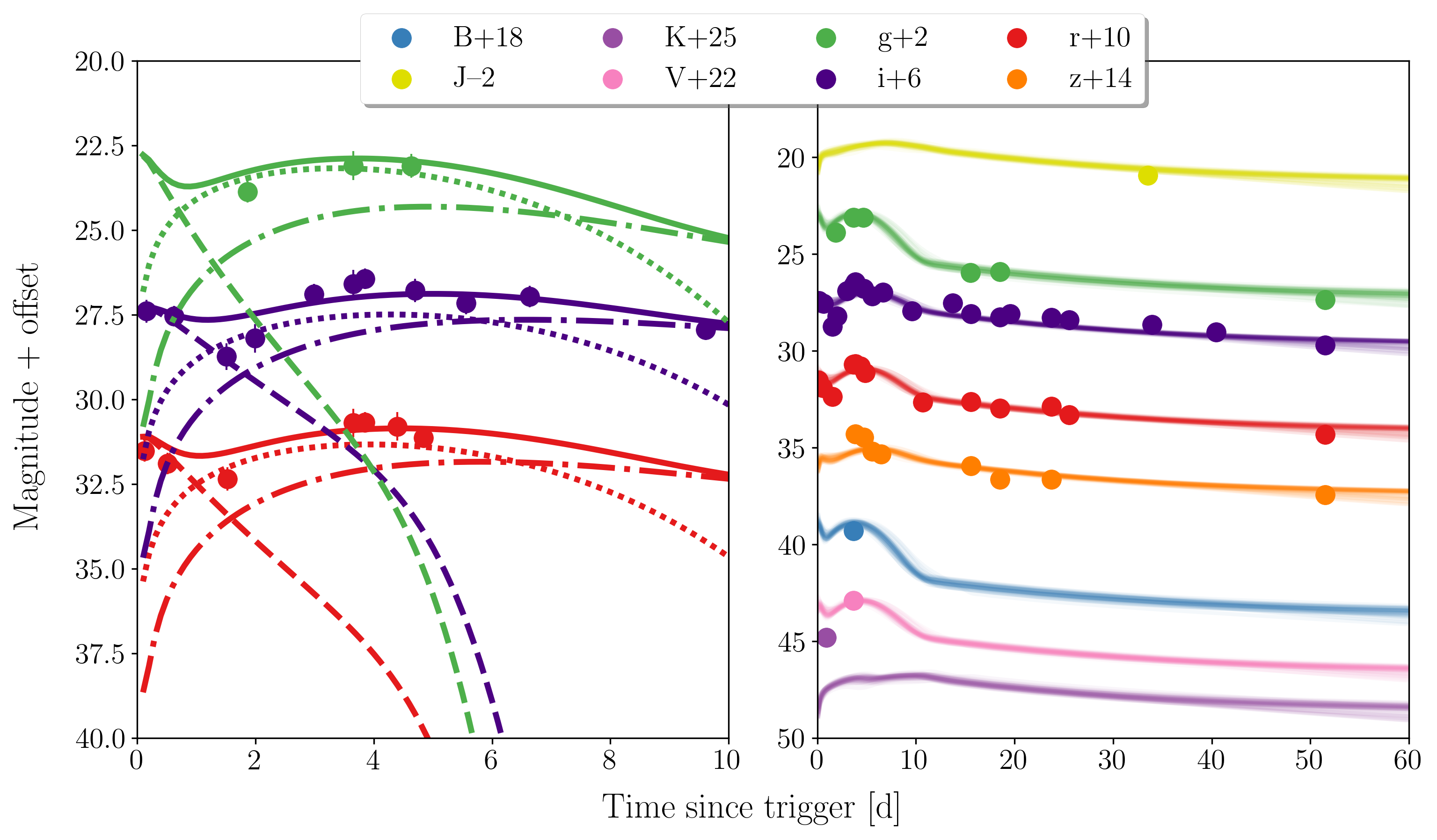}
    \caption{Light curve fits for the optical data of EP240414a. In the left panel, we show the first 10 days of observations, highlighting the best-fit model for the cocoon emission (dashed), CSM-interaction (dotted curves), and SN (dashed dotted curves) for \textit{g}-, \textit{r}-, and \textit{i}-band observations in green, red, and purple, respectively, with the total light as the solid curve. On the right hand panel we show the superposition of 50 random draws from the posterior for the full model for all our observations up to $\sim$~60~days.}
    \label{fig:lc_fit}
\end{figure*}

The observations of the first peak point towards a significant stellar envelope shocked by a relativistic jet, and the late-time observations indicate the presence of Ni-powered SN ejecta. This naturally suggests the presence of an additional emission component; the SN ejecta can interact with and shock previously ejected stellar material which will subsequently cool. Such interaction and cooling models have been used to explain observations of LFBOTs~\citep{Margalit2022, Ho2023}, which supports the broad consistency of the light curve at this epoch with the prototypical LFBOT, AT2018cow, although it should be noted that the spectrum of EP240414a at this peak is somewhat redder. This would suggest that the progenitor of this system is broadly similar to other long GRBs, but resides inside a dense shell of material, either ejected in the form of a dense stellar wind, eruptive mass-loss, or perhaps a common envelope. The first two peaks are then due to cocoon-emission (first peak) and the SN-CSM interaction (second peak), with the radioactive decay of the SN emerging later on. The lack of $\gamma$-ray detection could then be due to viewing angle effects, with the X-rays either the afterglow of the relativistic jet or from the shocked-jet~\citep{Nakar2017}. Interestingly, a bump in the optical on this timescale has already been seen in other SN~Ic-BLe such as SN2020bvc~\citep{2020Hobvc} and other SNe~\citep{Nakar2014} and is also interpreted to be powered by CSM-interaction. EP240414a may represent a more `extreme' version of such events with a stronger jet that then shows an earlier jet-CSM interaction signature that is otherwise too faint or missed in other events.

\subsection{Light curve modelling of EP240414a}
To explore jet and CSM interaction scenarios in more detail, we fit our optical observations with a combined model including emission from the `cocoon', the signature of the supernova ejecta interacting with the CSM, and radioactive decay of $^{56}$Ni. In particular, we follow the analytical model for cooling emission from stellar material shocked by a relativistic jet following~\citet{Nakar2017} and \citet{Piro2018} assuming that the shocked-stellar material is confined to an opening angle, $\theta_{\rm cocoon}$. We note that this model does not capture the emission from the shocked-jet itself, which is highly sensitive to the degree of mixing between the jet and the stellar-material and the shocked-jets velocity~\citep{Nakar2017}. 
For CSM-interaction, we use the one-zone model following~\citet{Margalit2022}, motivated by the success of this model to match the broad properties of the prototypical LFBOTs, such as AT2018cow. This model assumes a one-zone shell of ejecta interacting with a CSM with a `top-hat' profile located at a distance $R_{0}$ away from the progenitor with a thickness of $\Delta R$. Meanwhile to model the emission from the radioactive decay of $^{56}$Ni in a supernova we use the tried and tested `Arnett' model~\citep{Arnett1982}. For the CSM-interaction and Arnett models we assume a constant grey opacity, $\kappa = 0.07~\mathrm{cm}^{2}\mathrm{g}^{-1}$, consistent with other analyses of SN Ic-BLe~\citep{2019Taddia} and further assume a blackbody spectral energy distribution. We emphasize that our modelling is in many ways simplified, ignoring the contribution of the shocked-jet likely relevant at $\lesssim$~1~day, and using a one-zone model with the inbuilt assumption of spherical symmetry for CSM-interaction. However, we expect these models to broadly capture the different features of the light curve. 

We fit our optical observations using the combined model described above implemented in {\sc Redback}~\citep{Sarin2024_redback}, with {\sc pymultinest} sampling~\citep{Feroz2009} via {\sc Bilby}~\citep{bilby}, using broad uniform priors\footnote{We use the default prior for each of these models as implemented in {\sc Redback} available at \url{https://github.com/nikhil-sarin/redback/tree/master/redback/priors}.} and a standard Gaussian likelihood. To account for uncertainties in data reduction and different filter definitions between the various telescopes for our observations, we assume an additional systematic error of 0.3 mag added in quadrature to the statistical error for all data points. 

In Fig.~\ref{fig:lc_fit}, we show our fit to the multi-band optical light curve of EP240414a. In particular, in the left panel, we show the light curve in three filters over the first 10 days, where we see the initial contribution from the cocoon (dotted curves), which fades by the second peak to be subdominant to the CSM-interaction (dashed curves). The radioactive decay from the SN starts to dominate the light curve past $\sim 10$ days. On the right hand panel, we show the fit with the full model for all our observations. In general, we see good agreement of the data with our combined model apart from discrepancies with the K-band observations and the early CSM-interaction dominated light curve, particularly in \textit{i}-band.

Our inferred parameters for the SN are similar to the median properties from observed samples of SN~Ic-BL~\citep{Taddia2019}. In particular, we infer an ejecta mass of $\approx 0.9~M_{\odot}$ with a total kinetic energy of $\approx 2\times10^{49}$~erg. Meanwhile the CSM-interaction model suggests a total CSM-mass of $\approx 0.6~M_{\odot}$, with the shocked CSM's internal energy constrained to $\approx 1.4 \times 10^{47}$~erg at a radius of $\approx 1\times10^{13}$~cm, again consistent with expectations~\citep{Piro2021, Margalit2022}. However, the estimated parameters for the cocoon emission are inconsistent with expectations, with a total effective shocked cocoon mass of $\approx 0.8~M_{\odot}$, with a total energy of $10^{53}$~erg. This discrepancy could be a product of the simplified model employed for the cocoon emission in our modelling above. However, it is more likely that these extreme derived properties are instead suggesting that the early part of the light curve is powered by more than just the cooling emission from the shocked-cocoon, such as non-negligible contributions from the relativistic jet (especially in K-band) and from the shocked-jet itself. We leave more detailed modeling efforts to future work with the full X-ray, optical, and radio data set.

\section{Conclusions}
EP240414a is an FXT without a GRB counterpart. The event is associated with the core collapse of a massive star resulting in a broad-line~Type~Ic~SNe, similar to those seen in long-GRBs, but in a peculiar environment. The event has a large offset from its host galaxy which is a luminous spiral galaxy at \textit{z}=0.401. Before the emergence of the SN emission in EP24014a, we observe two prior peaks in the light curve. The first peak is modestly fading. The second peak has a fast rise time, and combined with the very blue spectroscopy taken at this time, EP240414a resembles observations of LFBOTs at a similar epoch. Our modelling of the light curve suggests an initially dominant contribution from cocoon emission, later dominated by CSM-interaction, finally followed by the SN radioactive decay becoming dominant at $\sim$10-15~days. Our observations associate the progenitors of typical long-GRBs to some FXTs even in the absence of a detected GRB, and find a possible connection between FXTs and LFBOTs, suggesting that these sources could all have similar origins.

\section*{Acknowledgments}
Based on observations collected at the European Southern Observatory under ESO programme(s): 113.26ET.002 PI Jonker, 113.26ET.008 PI Jonker, 110.24CF.022 1110.A-4348(V) Tanvir PI, 111.259Q.001 Jonker PI, and 113.26ET.004 PI Jonker; and GTC, under the International Time Programme of the CCI (International Scientific Committee of the Observatorios de Canarias of the IAC), operated on the island of La Palma by the Roque de los Muchachos under programm ID ITP23 PIs Jonker, Torres); and based on observations made in part with ALFOSC, which is provided by the Instituto de Astrofisica de Andalucia (IAA) under a joint agreement with the University of Copenhagen and the Nordic Optical Telescope, owned in collaboration by the University of Turku and Aarhus University, and operated jointly by Aarhus University, the University of Turku and the University of Oslo, representing Denmark, Finland and Norway, the University of Iceland and Stockholm University at the Observatorio del Roque de los Muchachos, La Palma, Spain, of the Instituto de Astrofisica de Canarias; and based in part on observations obtained at the Southern Astrophysical Research (SOAR) telescope, which is a joint project of the Minist\'{e}rio da Ci\^{e}ncia, Tecnologia e Inova\c{c}\~{o}es (MCTI/LNA) do Brasil, the US National Science Foundation’s NOIRLab, the University of North Carolina at Chapel Hill (UNC), and Michigan State University (MSU); and based in part on observations collected at the European Organisation for Astronomical Research in the Southern Hemisphere, Chile as part of PESSTO, (the Public ESO Spectroscopic Survey for Transient Objects Survey) ESO program 188.D-3003, 191.D-0935, 197.D-1075. The scientific results reported in this article are based in part on observations made by the {\em Chandra} X-ray Observatory under program number 29420, PI Jonker. This work made use of data supplied by the UK {\em Swift} Science Data Centre at the University of Leicester.

JNDvD, PGJ, JQV, and APCvH are supported by the European Union (ERC, StarStruck, 101095973). Views and opinions expressed are however those of the author(s) only and do not necessarily reflect those of the European Union or the European Research Council. Neither the European Union nor the granting authority can be held responsible for them. JQV additionally acknowledges support by the IAU-Gruber fundation.

DMS and MAPT acknowledge support by the Spanish Ministry of Science via the Plan de Generacion de conocimiento PID2021-124879NB-I00.

FEB acknowledges support from ANID-Chile BASAL CATA FB210003, FONDECYT Regular 1241005, and Millennium Science Initiative, AIM23-0001 and ICN12\_009.

JPA's work was funded by ANID, Millennium Science Initiative, ICN12\_009.

LG acknowledges financial support from AGAUR, CSIC, MCIN and AEI 10.13039/501100011033 under projects PID2023-151307NB-I00, PIE 20215AT016, CEX2020-001058-M, and 2021-SGR-01270.

MER acknowledges support from the research programme Athena with project number 184.034.002, which is financed by the Dutch Research Council (NWO).

POB acknowledges support from the UK Science and Technology Facilities Council through grant ST/W000857/1.

TEMB acknowledges financial support from the Spanish Ministerio de Ciencia e Innovaci\'on (MCIN), the Agencia Estatal de Investigaci\'on (AEI) 10.13039/501100011033, and the European Union Next Generation EU/PRTR funds under the 2021 Juan de la Cierva program FJC2021-047124-I and the PID2023-151307NB-I00 SNNEXT project, from Centro Superior de Investigaciones Cient\'ificas (CSIC) under the PIE project 20215AT016, and the program Unidad de Excelencia Mar\'ia de Maeztu CEX2020-001058-M.

NS acknowledges support from the Knut and Alice Wallenberg foundation through the “Gravity Meets Light” project.

S. Schulze is partially supported by LBNL Subcontract 7707915.  

\facilities{Chandra, GTC(EMIR, HiPERCAM, OSIRIS+), NOT(ALFOSC), NTT(EFOSC2), SOAR(Goodman-Red~Camera), Swift, VLT(FORS2,~MUSE,~X-shooter)}
\software{\texttt{Astropy} \citep{astropy:2013, astropy:2018, astropy:2022},
\texttt{Bilby}~\citep{bilby},
\texttt{CIAO}~\citep{2006Ciao},
\texttt{ESOReflex}~\citep{2013Freundling}, 
\texttt{ESORex}~\citep{ESOREX2015},
\texttt{Gelato}~\citep{2008Harutyunyan}, 
\texttt{LACosmic}~\citep{vanDokkum2001}, 
\texttt{MPDAF}~\citep{MPDAF2016}
\texttt{pPXF}~\citep{2017Cappellari},
\texttt{pymultinest}~\citep{Feroz2009},
\texttt{PyRAF}~\citep{Pyraf2012}, 
\texttt{Redback}~\citep{Sarin2024_redback},
\texttt{SNID}~\citep{2007Blondin},
\texttt{Vorbin}~\citep{2003Cappellari}
\texttt{XSPEC}~\citep{1996Arnaud}, 
\texttt{ZAP}~\citep{ZAP2015} 
} 

\clearpage

\appendix

\section{Tables of observations}
We include tables of the photometry, spectroscopy and X-ray observations of the source and the host galaxy that were obtained as part of this work as well as the publicly available photometry and x-ray detections. Each table contains the telescope and instrument, the time of the observation, the day since the EP trigger, the exposure time, and the filter with the addition of the AB magnitude for all photometry and references for the publicly available data. 

Table \ref{tab:photometry} contains photometry obtained with NOT, GTC, VLT, NTT, and SOAR between approximately 0.5 to 80 days post-burst obatined by this work. We use PanSTARRS and 2MASS standards to calibrate the photometry and obtain the AB magnitudes. Additionally, Table \ref{tab:photometrypublic} lists the early photometry provided through the General Coordinates Network (GCN) within the first few days after the burst.

Table \ref{tab:photometrypublic} lists the public data from GCN circulars which covered the first days of the event up to $\sim$10 days. Table \ref{tab:spectroscopy} gives a list of the spectra obtained with GTC and VLT with the first spectrum at 0.62 days and the final spectrum at 80.62 days. These spectra approximately cover the wavelengths between 4000 and 9000 \AA.

Table \ref{tab:x-rays} provides the publicly available X-ray data and the X-ray observations taken as part of this work by \textit{Chandra} and \textit{Swift}. 

Table \ref{tab:host} contains the host photometry obtained from our own images or publicly available images of the field.

\begin{table*}[ht!]
    \centering
    \caption{Photometry obtained with various ground-based telescopes for this work. UL in the AB magnitude column stands for the 1$\sigma$ upper limit.}
    \resizebox{0.9\textwidth}{!}{
    \begin{tabular}{ccccccc}
        Telescope & Instrument & Epoch (UT) & Since trigger (d) & Exp. time (s) & Filter & AB Magnitude \\ \hline
        NOT & ALFOSC & 2024-04-14 22:07:38 & 0.51 & $5\times300$ & r & 21.9$\pm$0.1 \\
        GTC & OSIRIS+ & 2024-04-15 00:46:43 & 0.62 & 150 & i & 21.55$\pm$0.03 \\
        NOT & ALFOSC & 2024-04-15 21:38:58 & 1.49 & $5\times300$ & g & 22 UL \\
        NOT & ALFOSC & 2024-04-15 22:06:34 & 1.51 & $4\times300$ & i & 22.74$\pm$0.26\\
        NOT & ALFOSC & 2024-04-15 22:28:40 & 1.53 & $4\times300$ & r & 22.35$\pm$0.16\\
        NTT & EFOSC2 & 2024-04-16 06:33:45 & 1.86 & $3\times250$ & g & 21.87$\pm$0.11\\
        NTT & EFOSC2 & 2024-04-16 06:38:39 & 1.87 & $3\times250$ & r & 21.6 UL\\
        NTT & EFOSC2 & 2024-04-16 06:43:33 & 1.87 & $3\times250$ & i & 21.4 UL\\
        VLT & MUSE & 2024-04-18 06:22:41 & 3.86 & 697 & r & 20.68$\pm$0.01\\
        VLT & MUSE & 2024-04-18 06:22:41 & 3.86 & 697 & i & 20.44$\pm$0.01\\
        VLT & MUSE & 2024-04-18 06:22:41 & 3.86 & 697 & z & 20.29$\pm$0.02\\
        Swift & UVOT & 2024-04-18 15:02:32 & 4.21 & 1833 & uvm2 & 21.75 $\pm$ 0.32 \\
        NOT & ALFOSC & 2024-04-19 01:06:42 & 4.64 & $2\times270$ & g & 21.11$\pm$0.18\\
        NOT & ALFOSC & 2024-04-19 01:14:05 & 4.64 & $2\times180$ & i & 20.79$\pm$0.17\\
        NOT & ALFOSC & 2024-04-19 02:31:49 & 4.66 & $3\times180$ & z & 20.47$\pm$0.08\\
        VLT & FORS2 & 2024-04-19 06:06:10 & 4.84 & 90 & r & 21.13$\pm$0.03 \\
        NOT & ALFOSC & 2024-04-19 23:12:14 & 5.56 & $4\times120$ & i & 21.16$\pm$0.14\\
        NOT & ALFOSC & 2024-04-19 23:22:33 & 5.56 & $6\times120$ & z & 21.19$\pm$0.40\\
        NOT & ALFOSC & 2024-04-20 20:38:12 & 6.45 & $9\times120$ & z & 21.33$\pm$0.27\\
        GTC & OSIRIS+ & 2024-04-21 01:11:46 & 6.64 & 150 & i & 20.97$\pm$0.10\\
        GTC & OSIRIS+ BB & 2024-04-24 00:34:00 & 9.61 & $9\times20$ & i & 21.94$\pm$0.10 \\
        VLT & X-shooter (acq) & 2024-04-25 02:04:31 & 10.68 & 5 & r & 22.65$\pm$0.20 \\
        SOAR & Red Camera & 2024-04-28 02:55:01 & 13.71 & $6\times500$ & i & 22.24$\pm$0.15 \\
        NOT & ALFOSC & 2024-04-29 23:04:11 & 15.55 & $3\times300$ & g & 23.97$\pm$0.40\\
        NOT & ALFOSC & 2024-04-29 23:12:51 & 15.56 & $2\times300$ & r & 22.63$\pm$0.15 \\
        NOT & ALFOSC & 2024-04-29 23:24:06 & 15.57 & $2\times300$ & i & 22.09$\pm$0.08 \\
        NOT & ALFOSC & 2024-04-29 23:35:24 & 15.58 & $5\times200$ & z & 21.94$\pm$0.29 \\
        GTC & HiPERCAM & 2024-05-02 21:54:46 & 18.50 & $16\times60$ & u & -- \\
        GTC & HiPERCAM & 2024-05-02 21:54:46 & 18.50 & $16\times60$ & g & 23.90$\pm$0.05 \\
        GTC & HiPERCAM & 2024-05-02 21:54:46 & 18.50 & $16\times60$ & r & 22.96$\pm$0.05 \\
        GTC & HiPERCAM & 2024-05-02 21:54:46 & 18.50 & $16\times60$ & i & 22.26$\pm$0.08\\
        GTC & HiPERCAM & 2024-05-02 21:54:46 & 18.50 & $16\times60$ & z & 22.63$\pm$0.06 \\
        GTC & OSIRIS+ & 2024-05-03 23:43:36 & 19.58 & 150 & i & 22.08$\pm$0.08 \\
        VLT & FORS2 & 2024-05-08 03:35:56 & 23.74 & $3\times50$ & r & 22.94$\pm$0.05 \\
        VLT & FORS2 & 2024-05-08 03:45:55 & 23.75 & $3\times100$ & g & 24.17$\pm$0.10\\
        VLT & FORS2 & 2024-05-08 03:55:10 & 23.75 & $3\times50$ & i & 22.25$\pm$0.06 \\
        VLT & FORS2 & 2024-05-08 04:03:01 & 23.76 & $5\times40$ & z & 22.60$\pm$0.15 \\
        NOT & ALFOSC & 2024-05-09 22:35:42 & 25.54 & $4\times300$ & r & 23.30$\pm$0.28 \\
        NOT & ALFOSC & 2024-05-09 22:57:49 & 25.55 & $4\times300$ & i & 22.41$\pm$0.13 \\
        GTC & EMIR & 2024-05-17 21:44:53 & 33.50 & 1440 & $K_s$ &  23.0 UL\\
        GTC & EMIR & 2024-05-17 22:40:39 & 33.54 & 1400 & J & 22.93$\pm$0.22 \\
        NOT & ALFOSC & 2024-05-17 22:18:07 & 33.95 & $9\times200$ & i & 22.65$\pm$0.14 \\
        NOT & ALFOSC & 2024-05-24 21:03:57 & 40.47 & $16\times200$ & i & 23.03$\pm$0.10 \\
        SOAR & Red Camera & 2024-05-28 00:40:23 & 43.62 & 2500 & i & 21.2 UL \\
        GTC & HiPERCAM & 2024-06-04 22:52:50 & 51.49 & $16\times60$ & u & -- \\
        GTC & HiPERCAM & 2024-06-04 22:52:50 & 51.49 & $16\times60$ & g & 25.35$\pm$0.13 \\
        GTC & HiPERCAM & 2024-06-04 22:52:50 & 51.49 & $16\times60$ & r & 24.31$\pm$0.06 \\
        GTC & HiPERCAM & 2024-06-04 22:52:50 & 51.49 & $16\times60$ & i & 23.7$\pm$0.06 \\
        GTC & HiPERCAM & 2024-06-04 22:52:50 & 51.49 & $16\times60$ & z & 23.44$\pm$0.08 \\
        VLT & MUSE & 2024-07-04 00:40:28 & 80.62 & 4$\times$697 & r & 24.66$\pm$0.16 \\
        VLT & MUSE & 2024-07-04 00:40:28 & 80.62 & 4$\times$697 & i & 24.01$\pm$0.13 \\
        VLT & MUSE & 2024-07-04 00:40:28 & 80.62 & 4$\times$697 & z & 23.71$\pm$0.16\\\hline
    \end{tabular}
    }
    \label{tab:photometry}
\end{table*}

\begin{table*}[ht!]
    \centering
    \caption{Spectroscopy obtained with various ground-based telescopes for this work.}
    \resizebox{0.8\textwidth}{!}{
    \begin{tabular}{cccccc}
        Telescope & Instrument & Epoch (UT) & Since trigger (d) & Exp. time (s) & Grism/arm \\ \hline
        GTC & OSIRIS+ & 2024-04-15 00:46:43 & 0.6225 & 1200 & R1000R \\
        VLT & MUSE & 2024-04-18 06:22:41 & 3.8558 & 4$\times$697 & $-$ \\
        VLT & FORS2 & 2024-04-19 06:26:14 & 4.8514 & 4$\times$600 & 300V \\
        VLT & X-shooter & 2024-04-25 02:07:02 & 10.6783 & 1200 & UVB \\
        VLT & X-shooter & 2024-04-25 02:07:07 & 10.6784 & 1200 & VIS \\
        VLT & X-shooter & 2024-04-25 02:07:11 & 10.6785 & 6$\times$300 & NIR \\
        GTC & OSIRIS+ & 2024-05-04 00:45:02 & 19.6214 & 1200 & R1000R \\
        VLT & MUSE & 2024-07-04 00:40:28 & 80.6182 & 4$\times$697 & $-$ \\
        \hline
    \end{tabular}
    }
    \label{tab:spectroscopy}
\end{table*}

\begin{table*}[ht!]
    \centering
    \caption{X-ray detections of EP240414a and the AGN in publicly available data used in this work. The EP and \textit{Swift} fluxes in this table are not corrected for the AGN contamination.}
    \resizebox{1.05\textwidth}{!}{%
    \begin{tabular}{cccccccc}
        Telescope      & Instrument & Mean epoch (UT)          & Since trigger (d) & Exp. time (ks) & Band (keV) & Flux (erg s$^{-1}$ cm$^{-2}$) & Reference \\\hline
        Einstein Probe  & WXT  & 2024-04-14 09:50:12  & Trigger & ---  & 0.5--4  & 3$\times$10$^{-9}$ &  \citet{Lian2024}  \\
        Einstein Probe  & FXT  & 2024-04-14 11:50:01  & 0.0757  & 7.2  & 0.5--10 & (3.5$\pm$0.8)$\times$10$^{-13}$ & \citet{Guan2024} \\
        \textit{Swift}  & XRT  & 2024-04-18 15:57     & 4.25   & 1.86 & 0.3--10 & 2.1$^{+0.9}_{-0.7}$$\times$10$^{-13}$ & This work \\
        \textit{Swift}  & XRT  & 2024-04-27 01:38     & 12.7   & 2.80 & 0.3--10 & $<$1.8$\times$10$^{-13}$ & This work \\
        \textit{Swift}  & XRT  & 2024-04-05 13:17     & 21.1   & 4.43 & 0.3--10 & 5.7$^{+0.4}_{-0.3}$$\times$10$^{-14}$ & This work \\
        \textit{Chandra} & ACIS & 2024-06-16 05:35:02  & 62.83  & 10.7 & 0.5--10 & Transient: $<$8$\times$10$^{-15}$  & \citet{Jonker2024} \\
         &   &   &   &  &  & AGN: 3$\times$10$^{-14}$ & \citet{Jonker2024} \\
         \hline
    \end{tabular}
    }%
    \label{tab:x-rays}%
\end{table*}

\begin{table*}[ht!]
    \centering
    \caption{Host galaxy photometry. Mag$_{2.1}$ is the absolute magnitude within a $2.1\arcsec$~radius aperture centred on the location of the galaxy, which includes only $\sim$2/3 of the light of the galaxy but avoids contamination by a nearby star. Mag$_{total}$ gives the magnitude as quoted in literature which might contain contamination from the nearby star.}
    \resizebox{0.6\textwidth}{!}{
    \begin{tabular}{ccccc}
        Telescope & Instrument & Filter & Mag$_{2.1}$ & 
        Mag$_\mathrm{total}$ \\
        \hline
        Swift & UVOT & UVM2 &21.65 $\pm$ 0.14  & 21.51 $\pm$ 0.09 \\
        GTC & HiPERCAM & $u$ & --- & ---\\
        SDSS & --- & $u$ & --- & 21.06 $\pm$ 0.21 \\
        VLT & FORS2 & $g$ & 20.46 $\pm$ 0.02 & ---\\
        SDSS & --- & $g$ & --- & 20.076 $\pm$ 0.03 \\
        VLT & FORS2 & $r$ & 19.44 $\pm$ 0.01 & ---\\
        SDSS & --- & $r$ & --- & 19.04 $\pm$ 0.02 \\
        VLT & FORS2 & $i$ & 18.98 $\pm$ 0.01 & --- \\
        SDSS & --- & $i$ & --- & 18.59 $\pm$ 0.02 \\
        VLT & FORS2 & $z$ & 18.66 $\pm$ 0.01 & ---\\
        SDSS & --- & $z$ & --- & 18.20 $\pm$ 0.05 \\
        VISTA & VIRCAM & $Y$ & 18.67 $\pm 0.07$ & ---\\
        GTC & EMIR & $J$ & 18.63 $\pm$ 0.01&---\\
        VISTA & VIRCAM & $H$ & 18.02 $\pm$ 0.05 & --- \\
        GTC & EMIR & $K$ & 18.03 $\pm$ 0.01  &--- \\
        WISE & --- & CH1 (3.4 $\mu$m) & --- & 17.58 $\pm$ 0.03 \\
        WISE & --- & CH2 (4.6 $\mu$m) & --- & 17.68 $\pm$ 0.05 \\
        WISE & --- & CH3 (12 $\mu$m) & --- & 16.83 $\pm$ 0.20 \\
        WISE & --- & CH4 (22 $\mu$m) & --- & 15.26 $\pm$ 0.58 \\
        \hline
    \end{tabular}
    }
    \label{tab:host}
\end{table*}

\begin{table*}[ht!]
    \centering
    \caption{Overview of the photometry that is publicly available for EP240414a obtained with various ground-based telescopes used in this work.}
    \resizebox{\textwidth}{!}{
    \begin{tabular}{cccccc}
        Telescope & Date (UT) & Since trigger (d)  & Filter & AB Magnitude & Reference \\ \hline
        LOT & 2024-04-14 & 0.13 & $r$ & 21.52 $\pm$ 0.12 & \citet{Aryan2024} \\
        LOT & 2024-04-14 & 0.15 & $i$ & 21.40 $\pm$ 0.16 & \citet{Aryan2024} \\
        Keck II (NIRES) & 2024-04-19 & 0.93 & $K'$ & 19.8 $\pm$ 0.1 & \citet{Karambelkar2024}\\
        Zeiss-2000 & 2024-04-15 & 1.56 & $R$ & 22.0 UL (3$\sigma$) & \citet{Belkin2024} \\
        Pan-STARRS1 & 2024-04-16 & 1.99 & $i$ & 22.2 $\pm$ 0.3 & \citet{Srivastav2024}\\
        Pan-STARRS1 & 2024-04-17 & 2.99 & $i$ & 20.9 $\pm$ 0.06 & \citet{Srivastav2024}\\
        LCO & 2024-04-18 & 3.66 & $g$ & $\sim$21.1 &\citet{Li2024}\\
        LCO & 2024-04-18 & 3.66 & $r$ & $\sim$20.7 &\citet{Li2024}\\
        LCO & 2024-04-18 & 3.66 & $i$ & $\sim$20.6 &\citet{Li2024}\\
        GMG & 2024-04-18 & 4.40 & $r$ & 20.8 $\pm$ 0.3 & \citet{Wang2024}\\
        Palomar 40-inch telescope (WINTER) & 2024-04-24 & 9.86 & $J$ & 19 UL& \citet{Karambelkar2024}\\\hline
    \end{tabular}
    }
    \label{tab:photometrypublic}
\end{table*}

\end{document}